\pdfoutput=1

\documentclass[a4paper]{amsart}

\usepackage{amsmath}
\usepackage{fix-cm}
\usepackage{graphicx}
\usepackage{hyperref}
\usepackage{amsthm}
\usepackage{amssymb}
\usepackage{bm}
\usepackage{verbatim}
\usepackage{inputenc}
\usepackage{cite}
\usepackage[a4paper,left=4cm,width=13cm,right=3cm,top=3cm,bottom=3cm]{geometry}

\usepackage[foot]{amsaddr}

\usepackage{caption}
\usepackage{subcaption}

\usepackage{epstopdf}

\begin{document}

\title[Collisionless equilibrium for the Force-Free Harris Sheet with low beta]{An exact collisionless equilibrium for the Force-Free Harris Sheet with low plasma beta}

\author{O. Allanson, T. Neukirch, F. Wilson and S. Troscheit}
\address{School of Mathematics and Statistics, University of St Andrews, St Andrews, KY16 9SS, UK}
\email{oliver.allanson@st-andrews.ac.uk}

\begin{abstract}
We present a first discussion and analysis of the physical properties of a new exact collisionless equilibrium for a one-dimensional nonlinear force-free magnetic field, namely the Force-Free Harris Sheet. The solution allows any value of the plasma beta, and crucially below unity, which previous nonlinear force-free collisionless equilibria could not. The distribution function involves infinite series of Hermite Polynomials in the canonical momenta, of which the important mathematical properties of convergence and non-negativity have recently been proven. Plots of the distribution function are presented for the plasma beta modestly below unity, and we compare the shape of the distribution function in two of the velocity directions to a Maxwellian distribution.
\end{abstract}

\date{\today}
\maketitle

\section{Introduction}
Equilibria are a suitable starting point for investigations of plasma instabilities and waves. Force-free equilibria, with fields defined by
\begin{eqnarray}
\nabla\cdot\mathbf{B}&=0,\nonumber\\
\nabla\times\mathbf{B}&=\mu_0\mathbf{j},\nonumber\\
\mathbf{j}\times\mathbf{B}&=\mathbf{0},  \label{eq:force-free}
\end{eqnarray}
are, for example, of particular relevance to the solar corona and other astrophysical plasmas, as well as the scrape-off layer in tokamaks; see for example refs. \cite{Wiegelmann}, \cite{Marshbook} and \cite{Fitzpatrick-2007} respectively. Equation (\ref{eq:force-free}) implies that the current density is everywhere-parallel to the magnetic field;
\begin{equation}
\mu_0\mathbf{j}=\alpha(\mathbf{r})\mathbf{B},
\end{equation}
or zero in the case of potential fields. If $\nabla\alpha\neq 0$ then the force-free field is nonlinear, whereas a constant $\alpha$ corresponds to a linear force-free field. Extensive discussions of force-free fields are given in refs. \cite{Marshbook} and \cite{Sakurai-1989}.

We consider one-dimensional (1D), non-relativistic collisionless plasmas, for which it is necessary to use kinetic theory. An equilibrium, characterised by the one-particle distribution function (DF) is a solution of the steady-state Vlasov equation \cite{Schindlerbook}. For a macroscopic equilibrium to be described, the DF must solve Maxwell's equations and describe force balance, via the coupling to the charge and current densities, as well as to the pressure. The difficulty of the problem in general lies in achieving self-consistency between the microscopic and macroscopic descriptions.

Current sheets are extremely important for reconnection studies, see ref. \cite{Priest-2000} for example. Three families of exact nonlinear force-free Vlasov-Maxwell equilibria are known \cite{Harrison-2009b,Wilson-2011,Abraham-Shrauner-2013, Allanson-2015a}, all of which describe 1D current sheets. The first family use the Force-Free Harris Sheet (FFHS) as their magnetic field profile \cite{Harrison-2009b,Wilson-2011}. The FFHS magnetic field is given by
\begin{equation}
\mathbf{B}=B_0\left(\text{tanh}\left(z/L\right), \text{sech}\left(z/L\right),0\right), \label{eq:FFHS}
\end{equation}
with $L$ the width of the current sheet, and $B_0$ the constant magnitude of the magnetic field. The second example uses Jacobi elliptic functions, of which the FFHS is a special case \cite{Abraham-Shrauner-2013}. The third example \cite{Allanson-2015a} will be discussed herein. We note work on `nearly' force-free equilibria \cite{Artemyev-2011}, with the FFHS modified by adding a small $B_z$ component. Examples of linear force-free VM equilibria \cite{Channell-1976, Sestero-1967, Bobrova-1979, Correa-Restrepo-1993, Bobrova-2001,Bobrova-2003} are discussed in ref. \cite{Harrison-2009a}. 

Two of the nonlinear force-free DFs known thus far \cite{Harrison-2009b, Wilson-2011, Abraham-Shrauner-2013} have the `drawback' of describing plasmas with a plasma-beta $(\beta_{pl})$ greater than one, due to the manner in which they were constructed. $\beta_{pl}$ is defined as the ratio of the thermal energy density to the magnetic energy density;
\begin{equation}
\beta_{pl}=\sum_s\beta_{pl,s}=\frac{2\mu_0k_B}{B_0^2}\sum_sn_sT_s, \label{eq:plasmabeta}
\end{equation} 
for $n_s$ and $T_s$ the number density and temperature - of species $s$ - respectively. In Cartesian geometry, and when gravity and electric fields are neglected, the fluid equation of motion\cite{Schindlerbook} becomes 
\begin{equation}
\frac{\partial}{\partial x_j}P_{ij}=(\mathbf{j}\times\mathbf{B})_i, \label{eq:gradp}
\end{equation}
for $P_{ij}$ the pressure tensor, and the Einstein summation convention applied over repeated indices. In this case, a $\beta_{pl}$ much less than one is typically taken to be consisitent with a force-free magnetic field. In the case of one `dynamical' component of the pressure tensor, $P$, force balance is described by $\nabla P=\mathbf{j}\times\mathbf{B}$ and $\beta_{pl}=2\mu_0P/B_0^2$. 

The basic theory of the technique used to reach the low plasma-beta regime, and the posing of the inverse problem are explained in Section II. Section III explains the inversion procedure used to find an equilibrium solution, with full detail in ref. \cite{Allanson-2015a}. In Section IV we present a discussion of some of the properties of the distribution function in the numerically accessible parameter regime. The first order moments of the DF are calculated in the Appendices, and used in Section IV to calculate the current sheet width. Finally we close with a summary and conclusions.

\section{Basic Theory}
An exact solution of the Vlasov equation is necessarily a function of the constants of motion \cite{Schindlerbook}. The equilibrium we shall consider varies only in one Cartesian spatial coordinate, namely $z$. This implies that the Hamiltonian, $H_s$, and the $x$ and $y$ canonical momenta, $p_{xs}$ and $p_{ys}$ for each particle species; 
\begin{eqnarray}
H_s&=&m_sv^2/2+q_s\phi,\nonumber\\
p_{xs}&=&m_sv_x+q_sA_x,\nonumber\\
p_{ys}&=&m_sv_y+q_sA_y,\nonumber
\end{eqnarray}
are conserved, with $q_s$ the charge of species $s$ and $\phi$ the scalar potential. $A_x$ and $A_y$ are components of the vector potential, with $A_z=0$ and $\mathbf{B}=\nabla\times\mathbf{A}$. The Vlasov equation can now be solved by any differentiable function $f_s(H_s,p_{xs},p_{ys})$, with the additional `physical' constraints being that $f_s$ is also normalisable, non-negative and has velocity moments of arbitrary order \cite{Schindlerbook}. The assumption of quasineutrality, $\sigma(z)=0$, implicitly defines the scalar potential as a function of the vector potential, $\phi_{qn}(A_x,A_y)$ \cite{Harrison-2009a}. For DFs of the form considered in this paper, see equation (11), the quasineutral scalar potential takes the form 
\begin{equation}
\phi_{qn}(A_x,A_y)=\frac{1}{e(\beta_e+\beta_i)}\ln (n_i/n_e).\label{eq:quasineutral}
\end{equation}
To be able to use the method of Channell \cite{Channell-1976} (described later), we choose our parameters such that strict neutrality ($n_i(A_x,A_y)=n_e(A_x,A_y)$) is satisfied. As already pointed out in ref. \cite{Harrison-2009a}, this implies that $\phi_{qn}=0$ due to equation (\ref{eq:quasineutral}). Our choice of parameters is mathematically equivalent to the condition used to derive micro-macroscopic parameter relationships, which will be discussed in Section III.

It has been shown in refs. \cite{Grad-1961, Bertotti-1963,Nicholson-1963, Sestero-1966,Harrison-2009a, Attico-1999} for example, that the 1D Vlasov-Maxwell equilibrium problem is analagous to that of a particle moving under the influence of a potential; with the relevant component of the pressure tensor, $P_{zz}$, taking the role of the potential; $(A_x,A_y)$ the role of position and $z$ the role of time. Under our assumptions this means that
\begin{eqnarray}
\frac{d^2A_x}{dz^2}=-\mu_0\frac{\partial P_{zz}}{\partial A_x},\label{eq:Amp1}\\
\frac{d^2A_y}{dz^2}=-\mu_0\frac{\partial P_{zz}}{\partial A_y}.\label{eq:Amp2}
\end{eqnarray}
These equations (or equivalent) are first seen in a complete sense in ref. \cite{Mynick-1979a}. Furthermore, the force-free equilibrium fields correspond to a  \emph{trajectory}, $(A_x(z),A_y(z))$, that is itself a \emph{contour};
\begin{equation} 
\frac{dP_{zz}}{dz}=0, \label{eq:Contour}
\end{equation}
of the \emph{potential}, $P_{zz}$ \cite{Harrison-2009b}. Equations (\ref{eq:Amp1})-(\ref{eq:Contour}) succinctly define the problem of calculating a 1D, quasineutral, force-free equilibrium. The difficulty lies in calculating the DF, $f_s$, given a macroscopic expression for $P_{zz}(A_x,A_y)$, formally defined
\begin{equation}
P_{zz}=\sum_sP_{zz,s}=\sum_sm_s\int {w_z}^2f_sd^3v,\label{eq:Pzz}
\end{equation} 
for $w_z=v_z-\langle v_z\rangle_s$, with $v_z$ the particle velocity and $\langle v_z\rangle_s$ the bulk velocity of species $s$ in the $z$ direction. Channell developed the theory of this problem\cite{Channell-1976}, with the added assumption of zero scalar potential from the offset, and a distribution function of the form
\begin{equation}
f_s=\frac{n_{0s}}{(\sqrt{2\pi}v_{th,s})^3}e^{-\beta_sH_s}g_s(p_{xs},p_{ys}).\label{eq:F_form}
\end{equation} 
The species dependent constants are $n_{0s}$, thermal velocity $v_{th,s}$ and $\beta_s=1/(k_BT_s)$. The `perturbation' to the Maxwellian, $g_s$, is an as yet unknown function of the canonical momenta. Note that the DF defined in equation (\ref{eq:F_form}) is an even function of $v_z$, giving no bulk flow in the $z$ direction, and hence $\langle v_z\rangle _s=0$. After integrating over $v_z$, the assumption of equation (\ref{eq:F_form}) in equation (\ref{eq:Pzz}) leads to 

\begin{eqnarray}
&&P_{zz}(A_x,A_y)=\frac{\beta_{e}+\beta_{i}}{\beta_{e}\beta_{i}}\frac{n_{0s}}{2\pi m_s^2v_{th,s}^2}\nonumber\\
&&\times\int_{-\infty}^\infty\int_{-\infty}^\infty e^{-\beta_{s}\left((p_{xs}-q_sA_x)^2+(p_{ys}-q_sA_y)^2\right)/(2m_s)}g_s(p_{xs},p_{ys})dp_{xs}dp_{ys}. \label{eq:Channell}
\end{eqnarray}

Equation (\ref{eq:Channell}) defines mathematically the inverse problem at hand: \emph{given a known macroscopic equilibrium, characterised by $P_{zz}(A_x,A_y)$, can we invert the integral equation to solve for $g_s(p_{xs},p_{ys})$?}

The inverse problem is not only non-unique regarding the form of the distribution function for a particular macroscopic equilibrium, but also for the form of $P_{zz}(A_x,A_y)$ for a given magnetic field. Given a specific magnetic field, i.e. a specific $\left(A_x,A_y\right)$, and a known $P_{zz}$ that satisfies equations (\ref{eq:Amp1})-(\ref{eq:Contour}), one can construct infinitely many new $\bar{P}_{zz}$ that also satisfy them;
\begin{equation}
\bar{P}_{zz}=\psi^\prime(P_{ff})^{-1}\psi(P_{zz}),\label{eq:Ptrans}
\end{equation} 
for differentiable and non-constant $\psi$, provided the LHS is positive (see ref. \cite{Harrison-2009a} for a discussion). These $\bar{P}_{zz}$  maintain a force-free equilibrium with the \emph{same magnetic field} as $P_{zz}$. The value of $P_{zz}$ evaluated on the force-free contour is $P_{ff}$. This paper takes the $P_{zz}$ used in refs. \cite{Harrison-2009b, Neukirch-2009, Wilson-2011}, and transforms it as in equation (\ref{eq:Ptrans}) with the exponential function according to 
\begin{equation}
\psi(P_{zz})=\exp\left[\frac{1}{P_0}\left(P_{zz}-P_{ff}\right)\right],\label{eq:Pfunc}
\end{equation}
with $P_0$ a free, positive constant. This gives $\bar{P}_{zz,ff}=P_0$, and so the plasma pressure can be as low or high as desired. Channell showed \cite{Channell-1976} that under the assumptions used in this paper,
\begin{equation}
P_{zz}(A_x,A_y)=\frac{\beta_e+\beta_i}{\beta_e\beta_i}n(A_x,A_y),\label{eq:Ppropn}
\end{equation}
where $n=n_i=n_e$. Equation (\ref{eq:plasmabeta}) then gives
\begin{equation}
\beta_{pl}=\frac{2\mu_0P_{zz,ff}}{B_0^2}=\frac{2\mu_0P_0}{B_0^2}.\nonumber
\end{equation} 
Hence, a freely chosen $P_0$ corresponds directly to a freely chosen $\beta_{pl}$. 

We note here that this pressure transformation can also be implicitly seen for the different linear force-free cases presented in the literature \cite{Channell-1976, Bobrova-2001, Sestero-1966,Attico-1999}, although this connection has never been made. The pressure function in ref. \cite{Sestero-1966} (and implicitly in ref. \cite{Bobrova-2001}) is an exponentiated version of that in refs. \cite{Channell-1976,Attico-1999}. A further interesting aspect is that the momentum dependent parts of the distribution functions are also related to each other exponentially.

\section{Calculating the distribution function}
The Harrison-Neukirch pressure function \cite{Harrison-2009b} is given by
\begin{equation}
P_{zz}=\frac{B_0^2}{2\mu_0}\left[\frac{1}{2}\text{cos}\left(\frac{2A_x}{B_0L}\right)+\text{exp}\left(\frac{2A_y}{B_0L}\right)+b\right], \label{eq:P_Harrison}
\end{equation}
with $b>1/2$ contributing to a `background' pressure sourced by a Maxwellian distribution, required for positivity. This is the pressure function that describes $\beta_{pl}>1$ regimes, and we are to transform according to equations (\ref{eq:Ptrans}) and (\ref{eq:Pfunc}) in order to realise $\beta_{pl}<1$, resulting in
\begin{equation}
\bar{P}_{zz}=P_0\exp\left\{\frac{1}{2\beta_{pl}}\left[\cos\left(\frac{2A_x}{B_0L}\right)+2\exp\left(\frac{2A_y}{B_0L}\right)              -1 \right]                     \right\}.\nonumber
\end{equation}
The $-1/(2\beta_{pl})$ term comes from results in ref. \cite{Harrison-2009b} regarding the $b>1/2$ term and the fact that ${P_{ff}=B_0^2/(2\mu_0)(1+(b-1/2))}$, readily seen for $z=0$, for example. Note that $P_{zz}$ is constant over $z$, and so we can evaluate at any $z$ to calculate $P_{ff}$. Exponentiation of $P_{zz}$ has clearly resulted in a complicated LHS of equation (\ref{eq:Channell}), and so the inverse problem defined above is mathematically challenging.  

Since exponentiation of the `summative' pressure function results in a `multiplicative' one, we shall exploit separation of variables by assuming $g_s\propto g_{1s}(p_{xs})g_{2s}(p_{ys})$, whilst noting that $\bar{P}_{zz}\propto \bar{P}_1(A_x)\bar{P}_2(A_y)$. This assumption leads to integral equations of the form
\begin{eqnarray}
\bar{P}_1(A_x)&\propto&\int_{-\infty}^{\infty}e^{-\beta_{s}\left(p_{xs}-q_sA_x\right)^2/(2m_s)}g_1(p_{xs})dp_{xs},\label{eq:P1tog1}\\
\bar{P}_2(A_y)&\propto&\int_{-\infty}^{\infty}e^{-\beta_{s}\left(p_{ys}-q_sA_y\right)^2/(2m_s)}g_2(p_{ys})dp_{ys},\label{eq:P2tog2}
\end{eqnarray}  
in which the LHS are formed of exponentiated cosine and exponential functions, respectively. These equations are 1D integral transforms, known primarily as Weierstrass transforms \cite{Widder-1951}, 
\begin{equation}
\mathcal{W}[f](x)=\frac{1}{\sqrt{4\pi}}\int_{-\infty}^\infty e^{-(x-y)^2/4}f(y)dy,\label{eq:Weier}
\end{equation}
used as Green's function solutions of the diffusion equation, see ref. \cite{Widder-1954} for example. One method of inversion \cite{Bilodeau-1962} involves the use of Hermite polynomials
\begin{equation}
H_n(x)=(-1)^ne^{x^2}\frac{d}{dx}e^{-x^2},\nonumber
\end{equation}
a complete orthogonal set for $f\in L^2(\mathbb{R},e^{-x^2}dx)$ \cite{Arfkenbook}. If one can expand the LHS of equation (\ref{eq:Weier}) in a Maclaurin series (with coefficients of expansion $a_n$) then the unknown function $f(y)$ can be written as 
\begin{equation}
f(y)=\sum_{n=0}^\infty a_nH_n(y/2).\nonumber
\end{equation}    
This is the method that we use to invert equations (\ref{eq:P1tog1}) and (\ref{eq:P2tog2}), to solve for $g_1(p_{1s})$ and $g_2(p_{ys})$.

In ref. \cite{Allanson-2015a}, this inversion procedure was performed, and we shall only outline the approach here. The first step is to Maclaurin expand the exponentiated pressure function of equation (\ref{eq:P_Harrison}) according to equations (\ref{eq:Ptrans}) and (\ref{eq:Pfunc}). Exponentiation of a power series is a combinatoric problem, and was tackled by E.T. Bell in ref. \cite{Bell-1934}. If $h(x)=\exp k(x)$ and 
\begin{equation}
k(x)=\sum_{n=1}^\infty \frac{1}{n!}\zeta_nx^n,\nonumber
\end{equation} 
then
\begin{equation}
h(x)=\sum_{n=0}^\infty \frac{1}{n!}Y_n(\zeta_1,\zeta_2,...,\zeta_n)x^n,\nonumber
\end{equation}
for $Y_n$ the $n$th Complete Bell Polynomial (CBP), and $Y_0=1$. These can be defined explicitly for $n\ge 1$ by Fa\`{a} di Bruno's determinant formula \cite{Johnson-2002} as the determinant of an $n\times n$ matrix;
 
\begin{equation}
Y_n(\zeta_1,\zeta_2,...\zeta_n)=\left|\begin{matrix}
  {n-1 \choose 0}\zeta_1 &{n-1 \choose 1}\zeta_2  &{n-1 \choose 2}\zeta_3  &\hdots &{n-1 \choose n-2}\zeta_{n-1} &{n-1 \choose n-1}\zeta_n \\
  -1 &{n-2 \choose 0}\zeta_1  & {n-2 \choose 1}\zeta_2 &\hdots &{n-2 \choose n-3}\zeta_{n-2} &{n-2 \choose n-1}\zeta_{n-1} \\
  0 & -1 & {n-3 \choose 0}\zeta_1  & \hdots&{n-3 \choose n-4}\zeta_{n-3} &{n-3 \choose n-3}\zeta_{n-2}\\
  \vdots  &\vdots &\vdots & &\vdots &\vdots\\
0& 0& 0& \hdots& {1\choose 0}\zeta_1&{1\choose 1}\zeta_2\\
0& 0& 0& \hdots& -1& {0\choose 0}\zeta_1\\
 \end{matrix}\right|.\label{eq:Bruno}
\end{equation}
For example $Y_1(\zeta_1)=\zeta_1$ and $Y_2(\zeta_1,\zeta_2)=\zeta_1^2+\zeta_2$. We include this determinant form here since this is the representation we use to plot the distribution function. Using these results, and a simple scaling argument in ref. \cite{Bell-1934}, the Maclaurin expansion of the transformed pressure is found to be
\begin{equation}
\bar{P}_{zz}=P_0e^{-1/(2\beta_{pl})}\sum_{m=0}^\infty a_{2m}\left(\frac{A_x}{B_0L}\right)^{2m}\sum_{n=0}^\infty b_n \left(\frac{A_y}{B_0L}\right)^n, \label{eq:Pmacl}
\end{equation} 
with
\begin{equation}
a_{2m}=e^{1/(2\beta_{pl})}\frac{(-1)^m2^{2m}}{(2m)!}Y_{2m}\left(0,\frac{1}{2\beta_{pl}}, 0 , ... , 0 , \frac{1}{2\beta_{pl}}\right),\label{eq:a2msimple}
\end{equation}
and
 \begin{equation}
b_n=e^{1/\beta_{pl}}\frac{2^n}{n!}Y_n\left(\frac{1}{\beta_{pl}}, ..., \frac{1}{\beta_{pl}}\right).\label{eq:bnsimple}
\end{equation}
This allows us to formally solve the inverse problem for the unknown functions $g_{1s}(p_{xs})$ and $g_{2s}(p_{ys})$ in terms of Hermite polynomials, giving
\begin{eqnarray}
&&f_s(H_s,p_{xs},p_{ys})=\displaystyle \frac{n_{0s}}{\left(\sqrt{2\pi}v_{th,s}\right)^3}\nonumber\\
&&\times\left[e^{-1/(2\beta_{pl})}\displaystyle\sum_{m=0}^\infty C_{2m,s}H_{2m}\left(\frac{p_{xs}}{\sqrt{2}m_sv_{th,s}}\right)\sum_{n=0}^\infty D_{ns}H_n\left(\frac{p_{ys}}{\sqrt{2}m_sv_{th,s}}\right)\right]e^{-\beta_sH_s},\label{eq:result}
\end{eqnarray}
for species-dependent and as yet unknown coefficients $C_{2m,s}$ and $D_{ns}$. The reason for this ambiguity is that the transforms defining our problem in equations (\ref{eq:P1tog1}) and (\ref{eq:P2tog2}) are not quite of the perfect form of the Weierstrass transform in equation (\ref{eq:Weier}). Since $P_{zz}$ is independent of species - see equation (\ref{eq:Pmacl}) - we have to ensure that taking the second moment of $f_i$ gives the same result as that of $f_e$, i.e. when computing the integral of equation (\ref{eq:Channell}). This is solved by fixing the parameters according to
\begin{eqnarray}
n_{0i}&=&n_{0e}=n_0,\hspace{3mm}P_0=n_0\frac{\beta_e+\beta_i}{\beta_e\beta_i}\nonumber\\
C_{2m,s}&=&\left(\frac{\delta_s}{\sqrt{2}}\right)^{2m}a_{2m},\hspace{3mm}D_{ns}=\text{sgn}(q_s)^n\left(\frac{\delta_s}{\sqrt{2}}\right)^{n}b_{n}.\nonumber
\end{eqnarray}
The dimensionless parameter $\delta_s>0$ is the species-dependent \emph{magnetisation parameter} \cite{Fitzpatrickbook}, also used as the fundamental ordering parameter in gyrokinetics \cite{Howes-2006};
\begin{equation}
\delta_s=\frac{m_sv_{th,s}}{|q_s|B_0L}\implies \frac{\delta_e}{\delta_i}=\sqrt{\frac{m_eT_e}{m_iT_i}}.
\end{equation}
It is the ratio of the thermal Larmor radius of species $s$, $\rho_s=v_{th,s}/\Omega_s$, for $\Omega_s=q_sB_0/m_s$, to the characteristic length scale of the system, $L$. When $\delta_s\ll 1$ then particle species $s$ is highly magnetised.

As yet, the distribution of equation (\ref{eq:result}), together with the micro-macroscopic conditions, is only a formal solution to the inverse problem posed. To be valid, it must be convergent, bounded and non-negative. We note here that infinite series in Hermite polynomials in velocity were used in Vlasov equilibrium studies in refs. \cite{Hewett-1976} and \cite{Suzuki-2008}, with the particular question of convergence raised in ref. \cite{Hewett-1976}. Convergence of our infinite series, as well as non-negativity and boundedness properties are proven in ref. \cite{Allanson-2015a}, and so will not be repeated here.

\section{Properties of the distribution function}
The nature of the inverse problem is to calculate a microscopic description of a system, given certain prescribed macroscopic data. Hence, one of the main tasks is to find the relationships between the characteristic parameters of each level of description. That is to say, given $(B_0,P_0,L)$ for example, what is their relation to $(m_s, q_s, v_{th,s}, n_{0s})$?
\subsection{Current sheet width}
Currently, there are six free parameters that will determine the nature of the equilibrium. These are $n_{0}$, $\beta_{pl}$, $\beta_{th,i}$, $\beta_{th,e}$,  $\delta_i$ and $\delta_e$. $n_{0}$ is in principle fixed by ensuring that the DF is normalised to the total particle number. As yet we have no information regarding the width of the current sheet $L$. To this end we shall consider bulk velocities $\langle v_x\rangle_s$ and $\langle v_y\rangle_s$, obtained from the first moment of the DF. The calculations in Appendices A and B, together with the fact that $B_0=\sqrt{2\mu_0P_0/\beta_{pl}}$ give
\begin{eqnarray*}
\langle v_x\rangle_{s}&=\displaystyle\sqrt{\frac{2}{\mu_0\beta_{pl}P_0}}\frac{1}{Lq_s\beta_s}\frac{\text{sinh}\left(\frac{z}{L}\right)}{\text{cosh}^2\left(\frac{z}{L}\right)},\\
\langle v_y\rangle_{s}&=\displaystyle\sqrt{\frac{2}{\mu_0\beta_{pl}P_0}}\frac{1}{Lq_s\beta_s}\frac{1}{\text{cosh}^2\left(\frac{z}{L}\right)}.
\end{eqnarray*}
We can identify the coefficient of the $z$ dependent profiles as the amplitude of the bulk velocities, $\langle v_x\rangle_{s}$ and $\langle v_y\rangle_{s}$, as $u_{s}$, given by
\begin{equation}
u_{s}=\sqrt{\frac{2}{\mu_0\beta_{pl}P_0}}\frac{1}{Lq_s\beta_s},\label{eq:us}
\end{equation}
giving
\begin{eqnarray}
(u_{i}-u_{e})^2&=&\frac{2}{\mu_0\beta_{pl}P_0L^2q^2}\left(\frac{\beta_e+\beta_i}{\beta_e\beta_i}\right)^2,\\
\implies L &=&\frac{1}{q}\sqrt{\frac{2(\beta_e+\beta_i)}{\mu_0n_0\beta_e\beta_i(u_{i}-u_{e})^2\beta_{pl}}},\label{eq:length1}
\end{eqnarray}
where $q=|q_s|$. Interestingly, this is almost identical to the expression found in ref. \cite{Neukirch-2009} for the current sheet width of the Harrison-Neukirch equilibrium, with the addition of the $\beta_{pl}^{1/2}$ factor in the denominator. It is readily seen that, given some fixed $B_0$, $L\propto\beta_{pl}^{-1/2}$. This makes sense in that, by raising the number density $n_0$, and hence $\beta_{pl}$, there are simply more current carriers available to produce $\mathbf{j}$, and hence the width $L$ can reduce. By manipulating equation (\ref{eq:us}) one can show that the amplitudes of the fluid velocities are given by 
\begin{equation}
\frac{u_{s}}{v_{th,s}}=2\text{sgn}(q_s)\frac{\delta_s }{\beta_{pl}}=2\text{sgn}(q_s)\frac{\rho_s}{L\beta_{pl}}.
\end{equation}
Once again, this is almost identical to the expression found in ref. \cite{Neukirch-2009}, with the addition of a $\beta_{pl}$ factor in the denominator. 
\begin{comment}
\subsection{Zero Temperature Limit}
It is interesting to consider the functional dependence of $g_s$ on temperature, and the zero temperature limit of the distribution in general. The zero temperature limit of the Hermite expanded $g_s$ functions have been calculated \cite{Allanson-2015a} and shown to exist in a purely mathematical sense, treating $\beta_{pl}$ as an independent parameter. The limit is readily seen to be
\begin{equation*}
\lim _{v_{th,s}^2\to 0^+}g_s(q_sp_x,q_sp_y)=\bar{P}_{zz}(p_x,p_y).
\end{equation*} 
Whether or not this limit is meaningful in a physical sense, depends on the interpretation of $\beta_{pl}$ at zero temperature. The instinctual response is to say that $\beta_{pl}$ should be zero at zero temperature since pressure goes to zero, giving a singularity for $g_s$ and $\bar{P}_{zz}$. The nature of the magnetic field at zero temperature should be in question however, and for that matter the width of the current sheet. This all calls for a rather subtle analysis, and perhaps could form future work. Certainly, as a mathematical tool, the vanishing-width limit of the Gaussian function as a Dirac Delta is particularly useful, as shown in ref. \cite{Allanson-2015a}. Note that the Harrison-Neukirch DF \cite{Harrison-2009b} displayed functional equality of the pressure and the unknown parts of the distribution for arbitrary finite temperatures, i.e. their $g_s$ function was an eigenfunction of the Weierstrass transform \cite{Wolf-1977}.
\end{comment}
\subsection{Plots of the distribution function}
\begin{figure}
\centering
\begin{subfigure}[b]{0.5\textwidth}
\includegraphics[width=\textwidth]{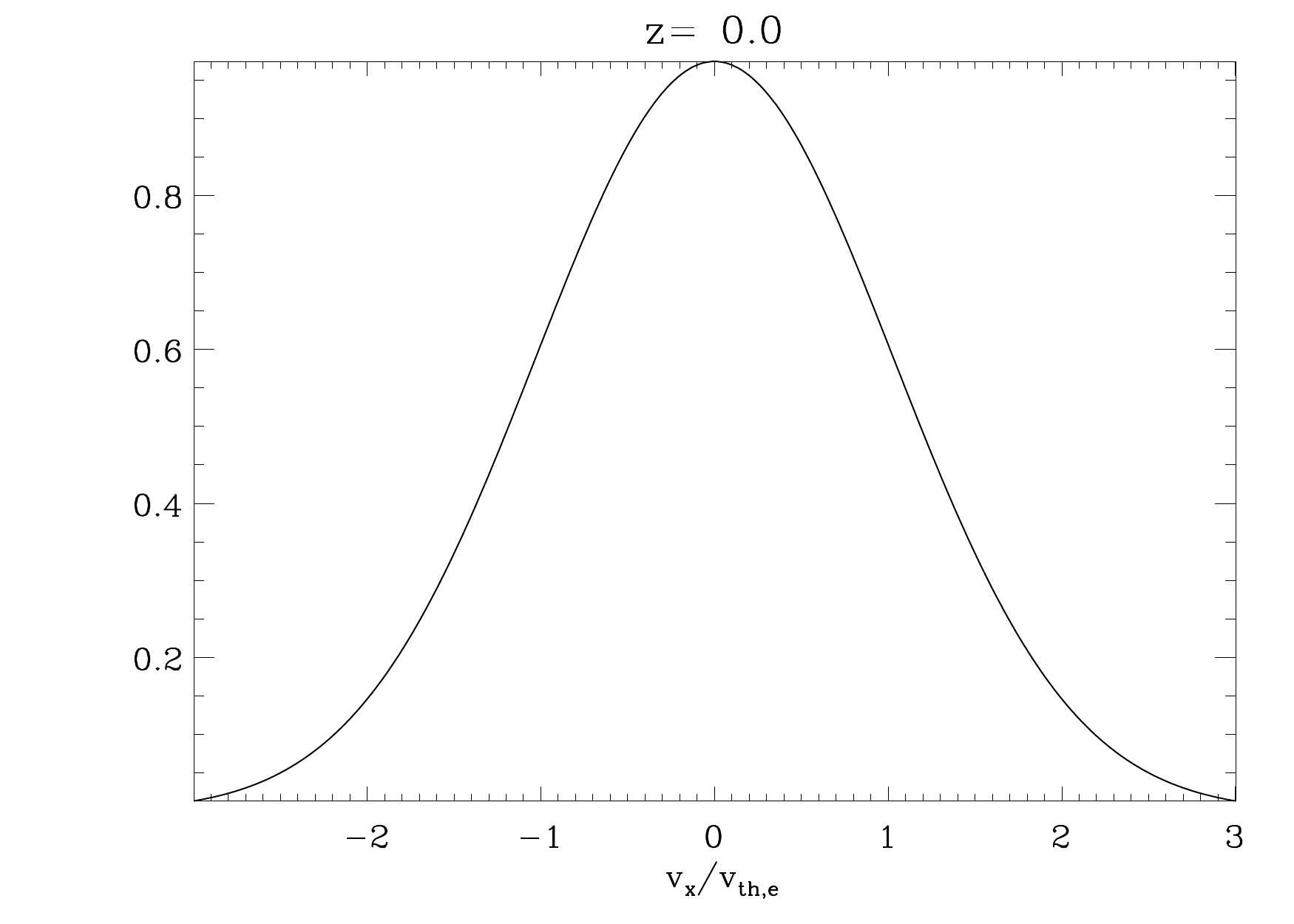}
 \caption{\label{fig:1a}}
\end{subfigure}
\begin{subfigure}[b]{0.5\textwidth}
\includegraphics[width=\textwidth]{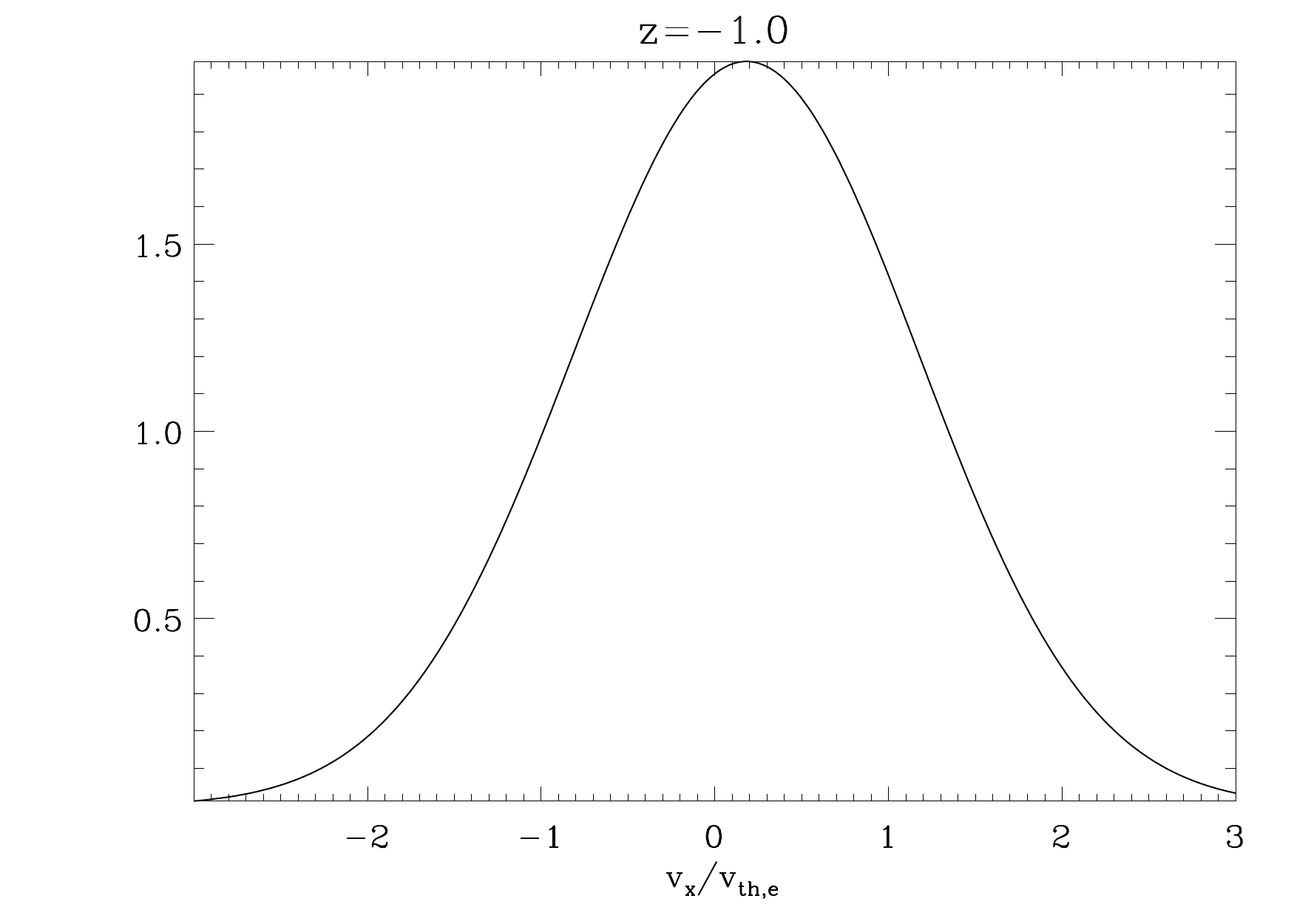}
 \caption{\label{fig:1b}}
\end{subfigure}
\begin{subfigure}[b]{0.5\textwidth}
\includegraphics[width=\textwidth]{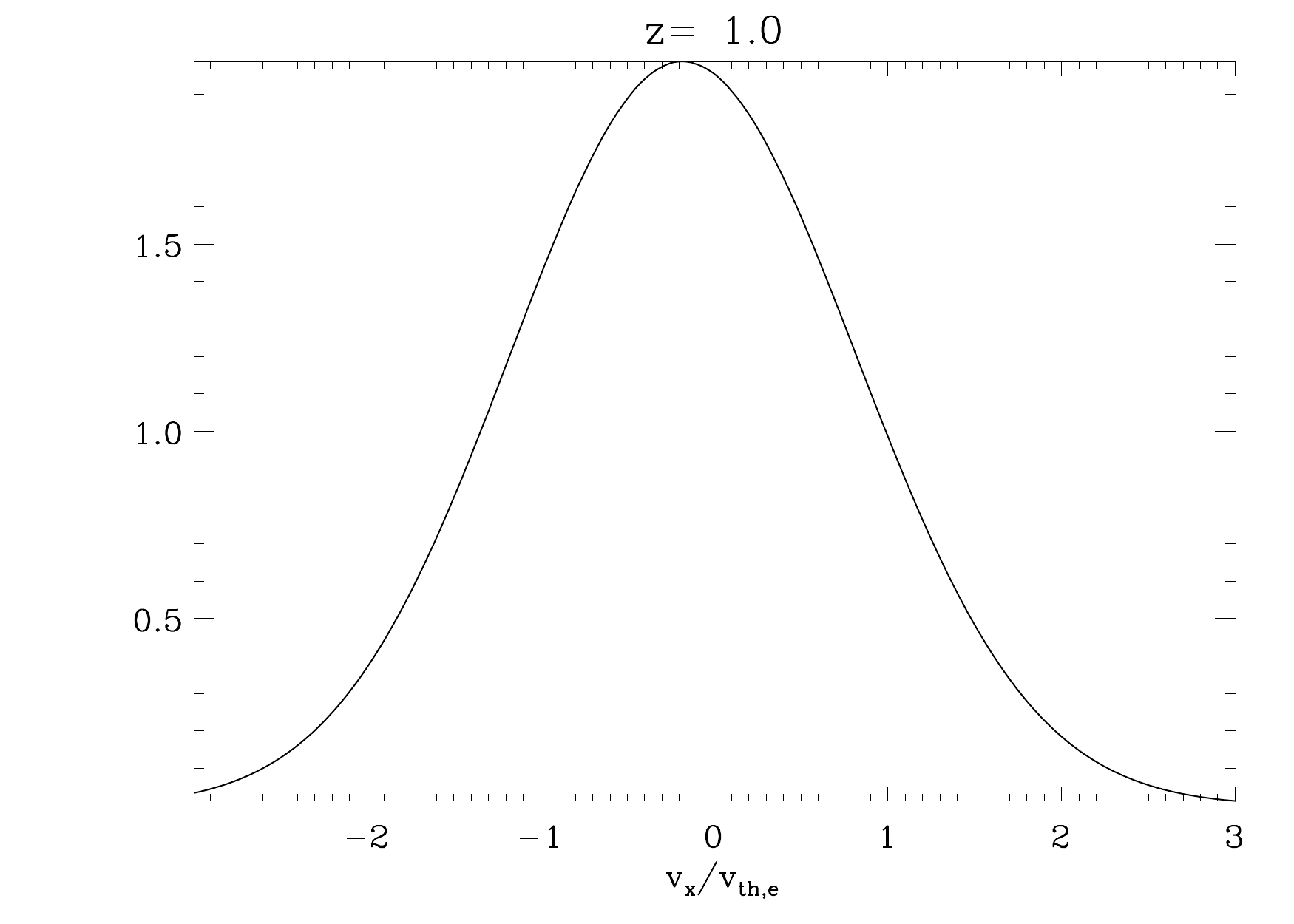}
 \caption{\label{fig:1c}}
\end{subfigure}
\caption{The $v_x$ variation of $f_e$ for $z/L=0$ (\ref{fig:1a}),  $z/L=-1$ (\ref{fig:1b}) and  $z/L=1$ (\ref{fig:1c}). $\beta_{pl}=0.85$ and $\delta_e=0.15$. Note the antisymmetry of the $z=\pm 1$ plots with respect to each other.}
 \end{figure}

Having found mathematical expressions for the DFs, we now present different plots of their dependence on $v_x$ and $v_y$, for $z/L=0,-1,1$. Plotting $f_s$ is a challenging numerical task, particularly for the low-$\beta_{pl}$ regime as when $\beta_{pl}<1/2$, the $C_{2m,s}$ are readily seen to be of the order
\begin{equation*}
\left(\frac{1}{\sqrt{2}}\right)^{2m}\frac{(-1)^m}{(2m)!}\left(\frac{\delta_s}{\beta_{pl}}\right)^{2m},
\end{equation*}
since $Y_{2m}$ is a polynomial of order $2m$ in $1/(2\beta_{pl})$. While it has been proven that the series' with which we represent the DFs are convergent for all values of the relevant parameters, attaining numerical convergence for the low-$\beta_{pl}$ regime, particularly for the $p_{xs}$ dependent sum is thus far proving difficult. Here we present plots for $\beta_{pl}=0.85$ and $\delta_i=\delta_e=0.15$.  As aforementioned we use Fa\`{a} di Bruno's determinant formula in equation (\ref{eq:Bruno}) to calculate the CBP's, and a recurrence relation for the Hermite Polynomials. We acknowledge that this $\beta_{pl}$ is only modestly below unity, however it represents a value of which we are confident of our numerics for both the $p_x$ and $p_y$ dependent sums. In figures (\ref{fig:1a})-(\ref{fig:1c}) we plot the $v_x$ variation of our electron distribution function, as a representative example (the $v_y$ plots are qualitatively similar). First of all we note that the DFs appear to have only a single maximum, and fall off as $v_x\to\pm \infty$. This is to be contrasted with the plots of the DF using the additive pressure \cite{Neukirch-2009}, which can have multiple peaks. Thus far we have not found any indication of multiple peaks in the parameter regime that we have been able to explore. However, this does not mean that multiple peaks can not appear, for example for lower values of the $\beta_{pl}$.

\begin{figure}
\centering
\begin{subfigure}[b]{0.5\textwidth}
\includegraphics[width=\textwidth]{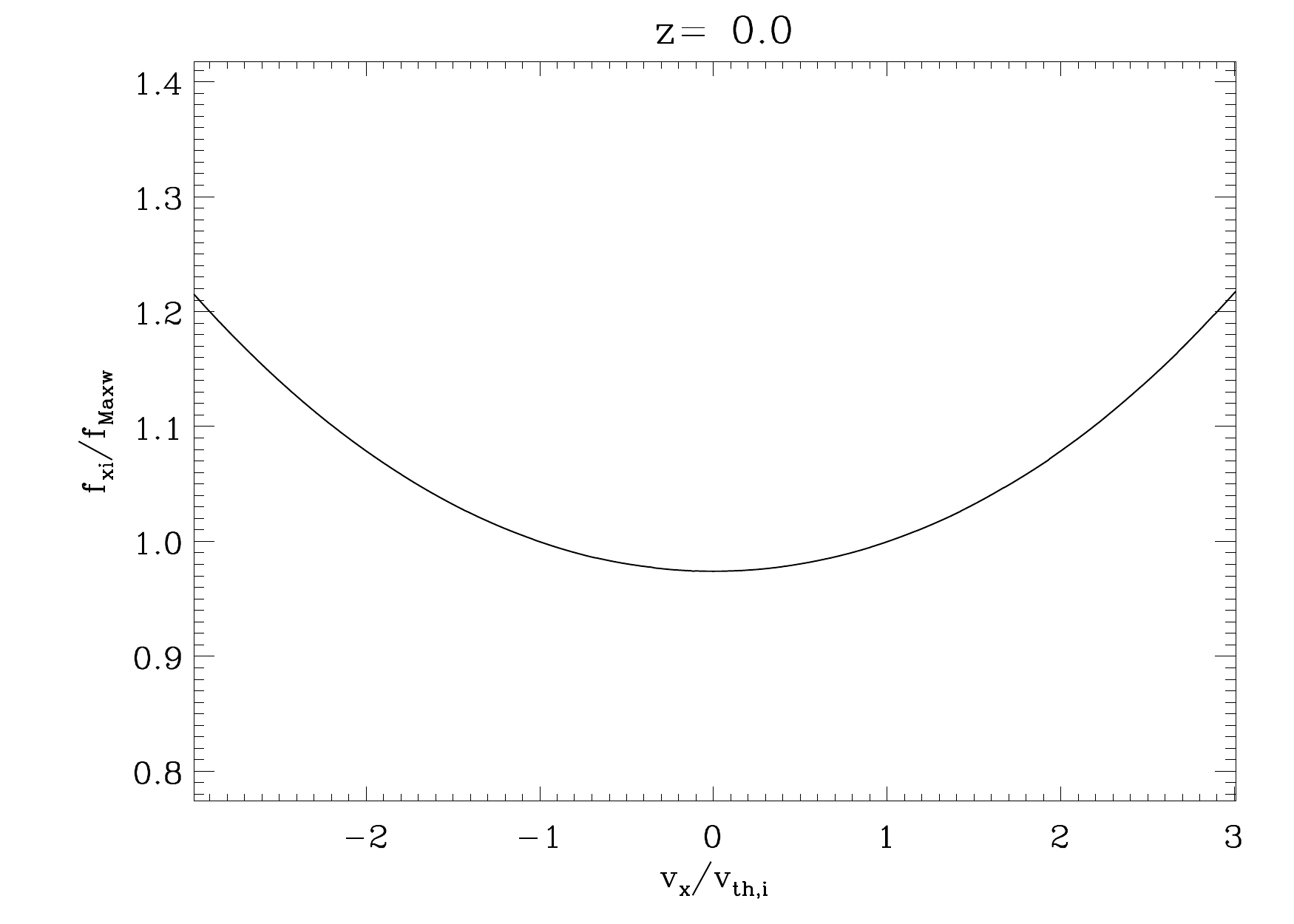}
 \caption{\label{fig:2a}}
\end{subfigure}
\begin{subfigure}[b]{0.5\textwidth}
\includegraphics[width=\textwidth]{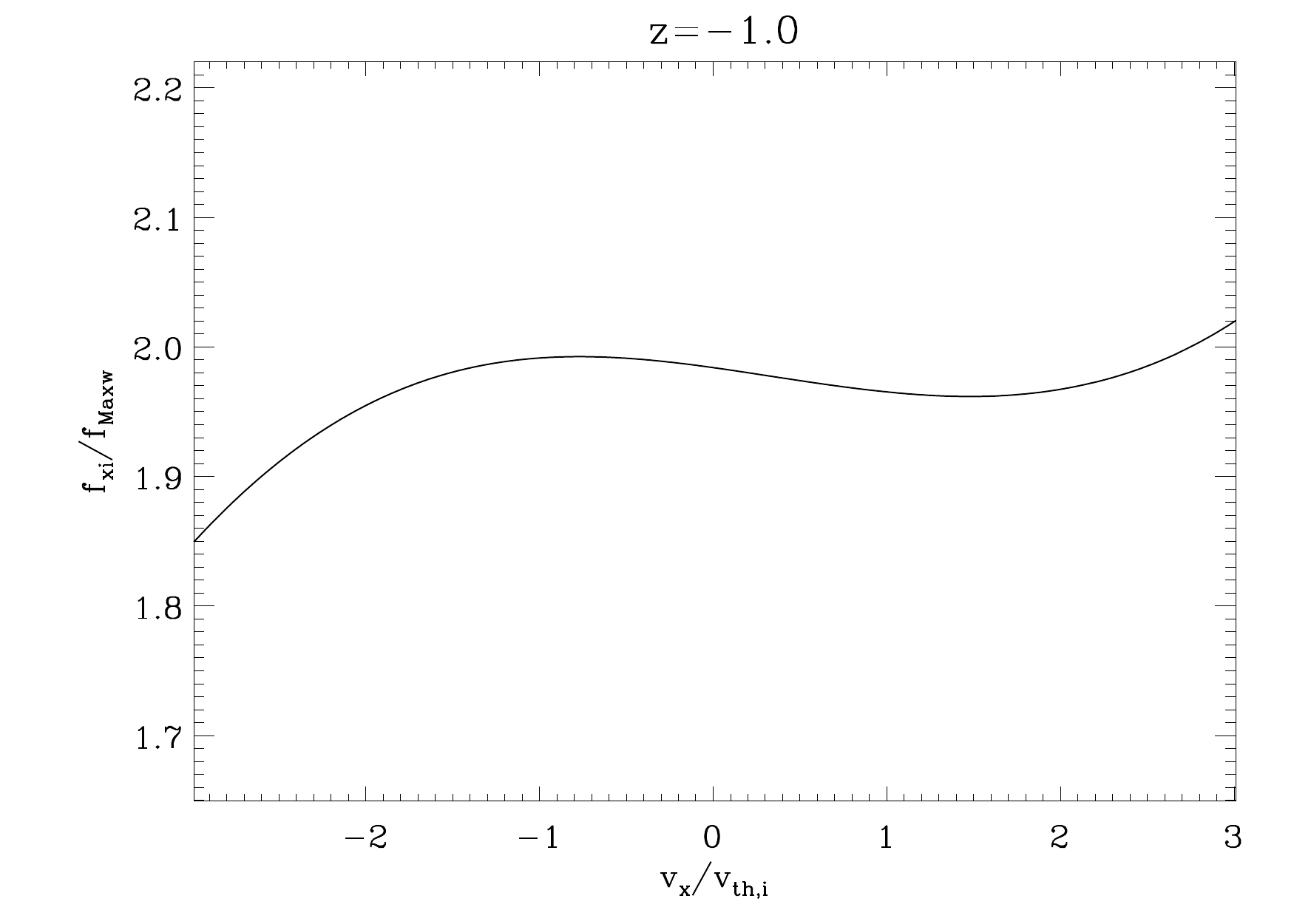}
 \caption{\label{fig:2b}}
\end{subfigure}
\begin{subfigure}[b]{0.5\textwidth}
\includegraphics[width=\textwidth]{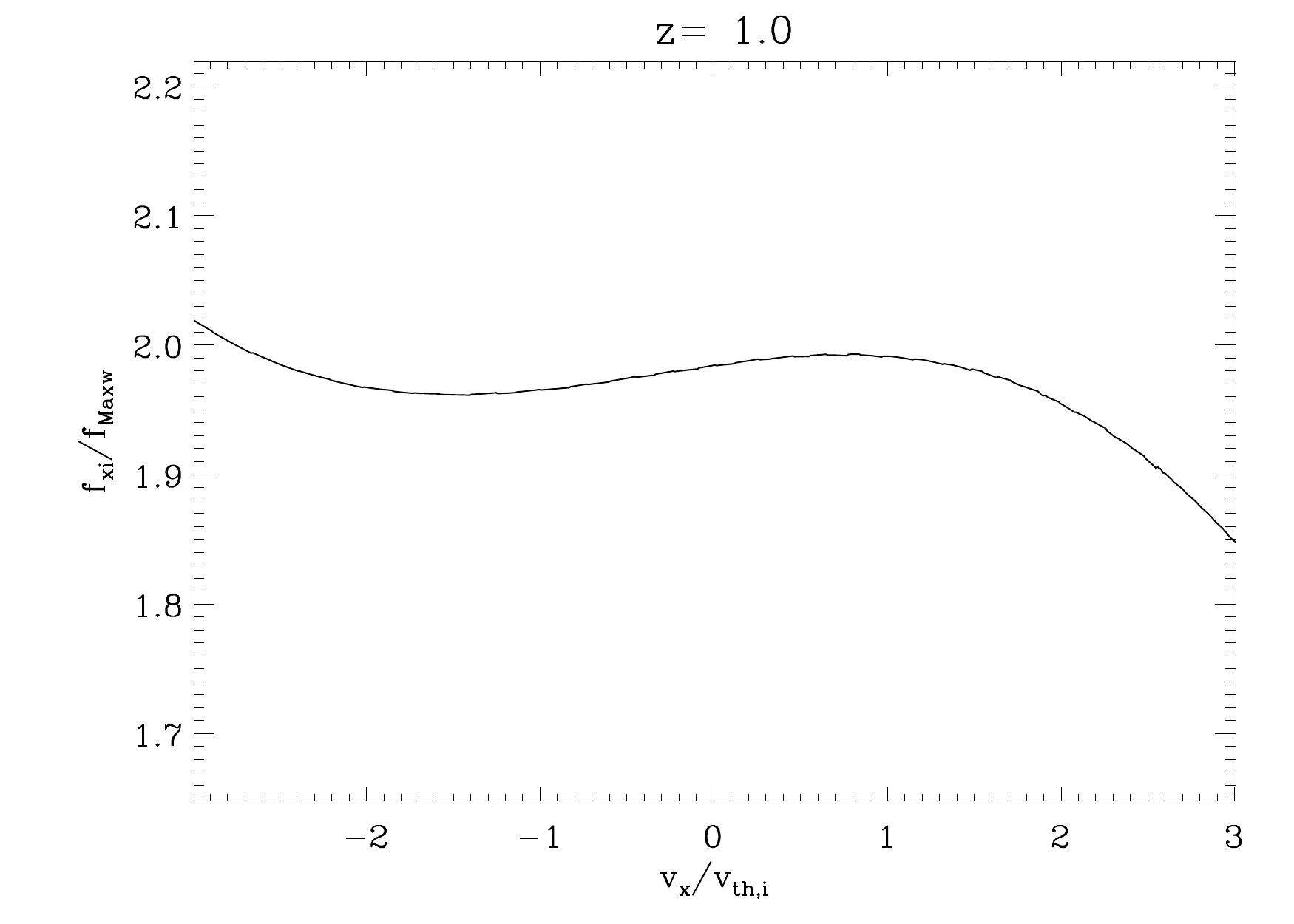}
 \caption{\label{fig:2c}}
\end{subfigure}
\caption{The $v_x$ variation of $f_i/f_{Maxw,i}$ for $z/L=0$ (\ref{fig:2a}),  $z/L=-1$ (\ref{fig:2b}) and  $z/L=1$ (\ref{fig:2c}). $\beta_{pl}=0.85$ and $\delta_i=0.15$. Note the antisymmetry of the $z=\pm 1$ plots with respect to each other.}
 \end{figure}

\begin{figure}
\centering
\begin{subfigure}[b]{0.5\textwidth}
\includegraphics[width=\textwidth]{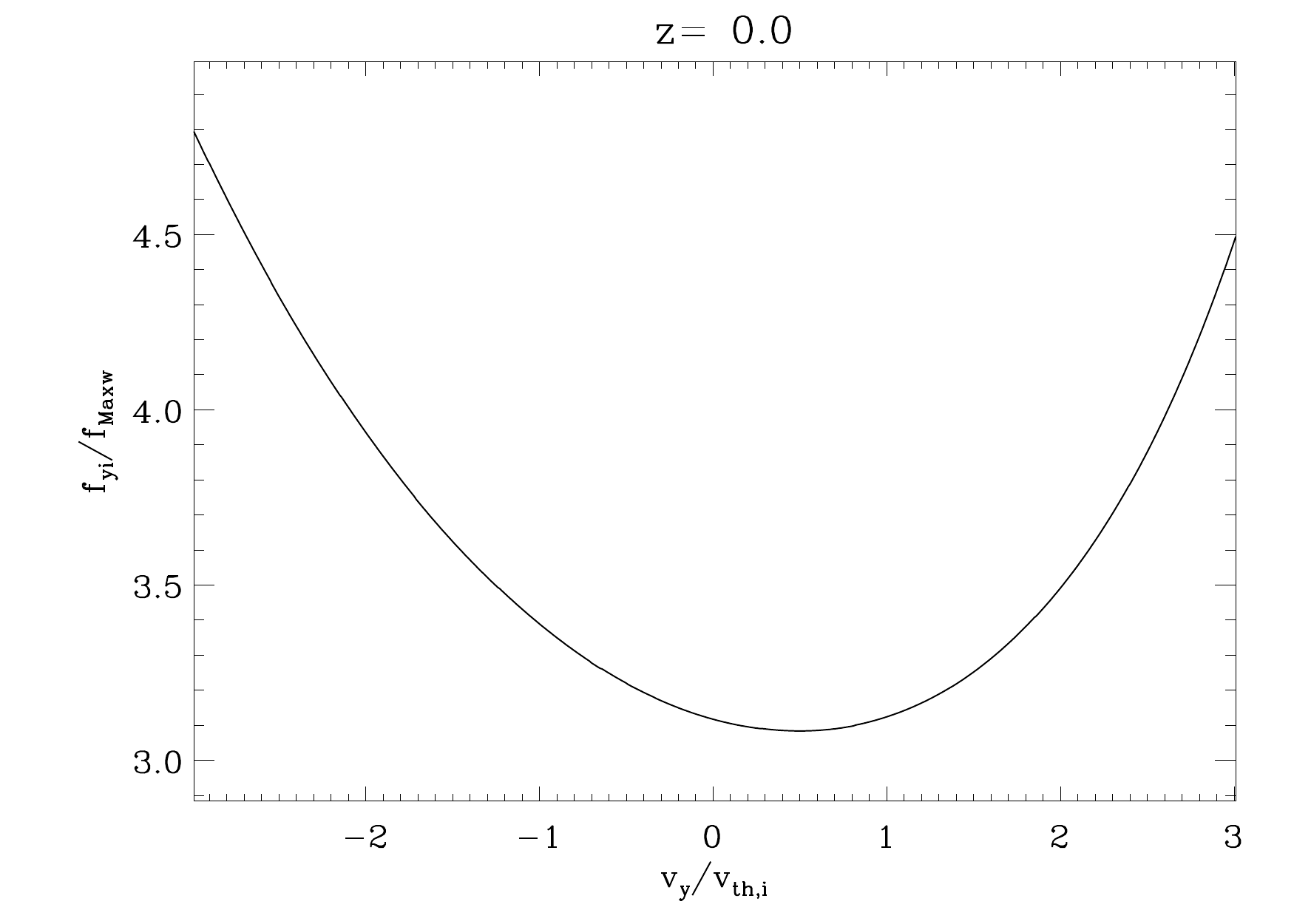}
 \caption{\label{fig:3a}}
\end{subfigure}
\begin{subfigure}[b]{0.5\textwidth}
\includegraphics[width=\textwidth]{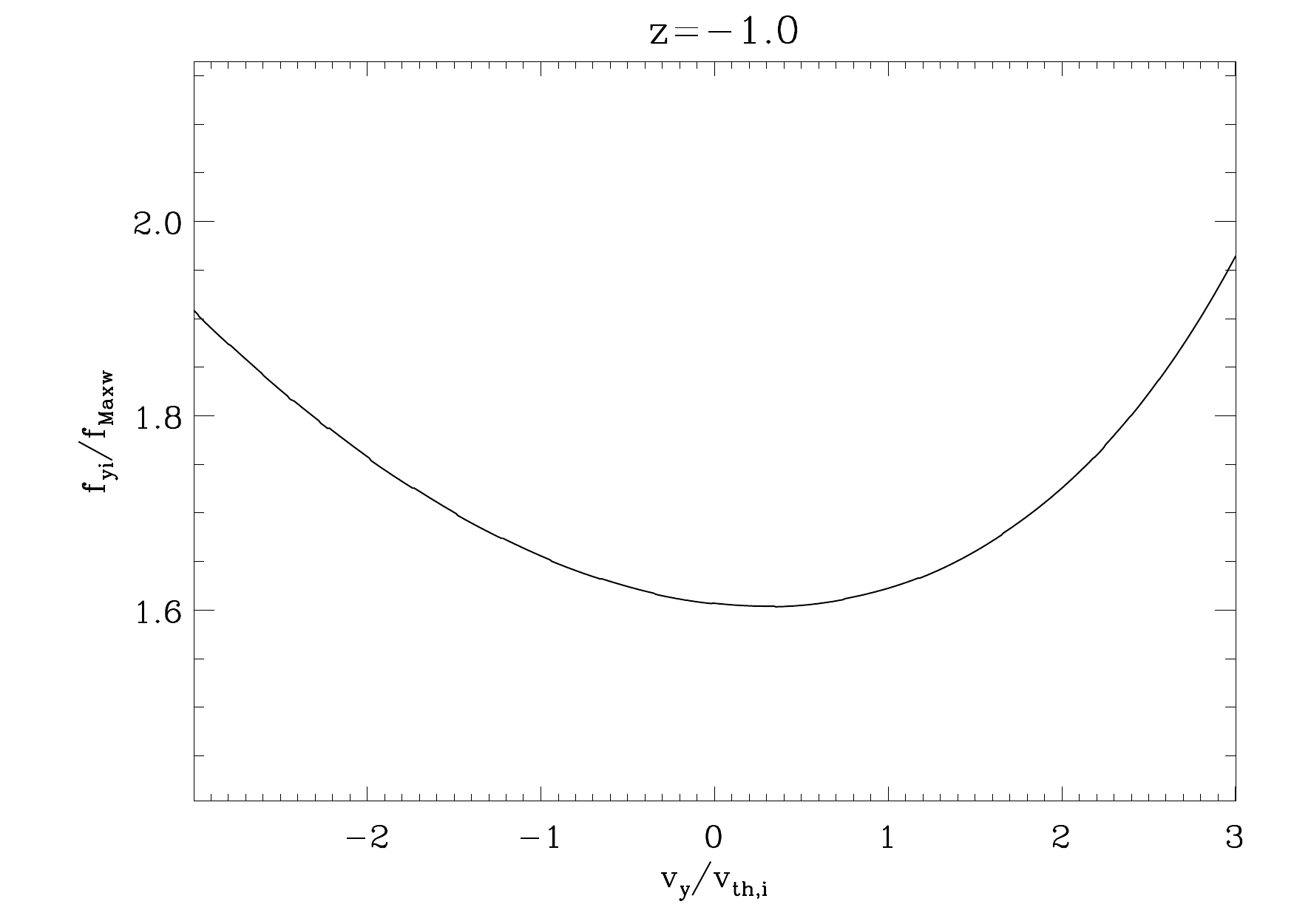}
 \caption{\label{fig:3b}}
\end{subfigure}
\begin{subfigure}[b]{0.5\textwidth}
\includegraphics[width=\textwidth]{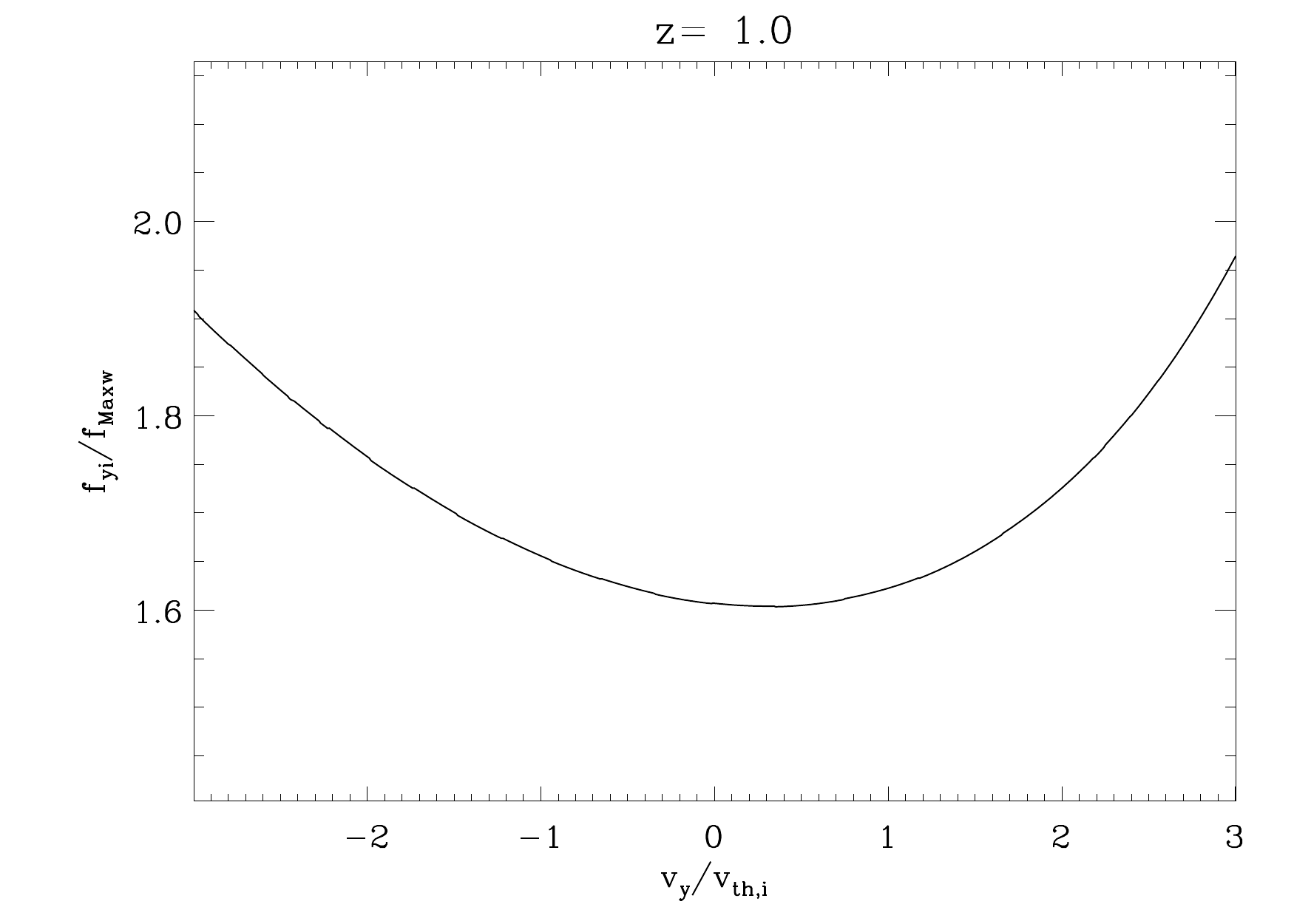}
 \caption{\label{fig:3c}}
\end{subfigure}
\caption{The $v_y$ variation of $f_i/f_{Maxw,i}$ for $z/L=0$ (\ref{fig:3a}),  $z/L=-1$ (\ref{fig:3b}) and  $z/L=1$ (\ref{fig:3c}). $\beta_{pl}=0.85$ and $\delta_i=0.15$. Note the symmetry of the $z=\pm 1$ plots with respect to each other.}
\end{figure}

\begin{figure}
\centering
\begin{subfigure}[b]{0.5\textwidth}
\includegraphics[width=\textwidth]{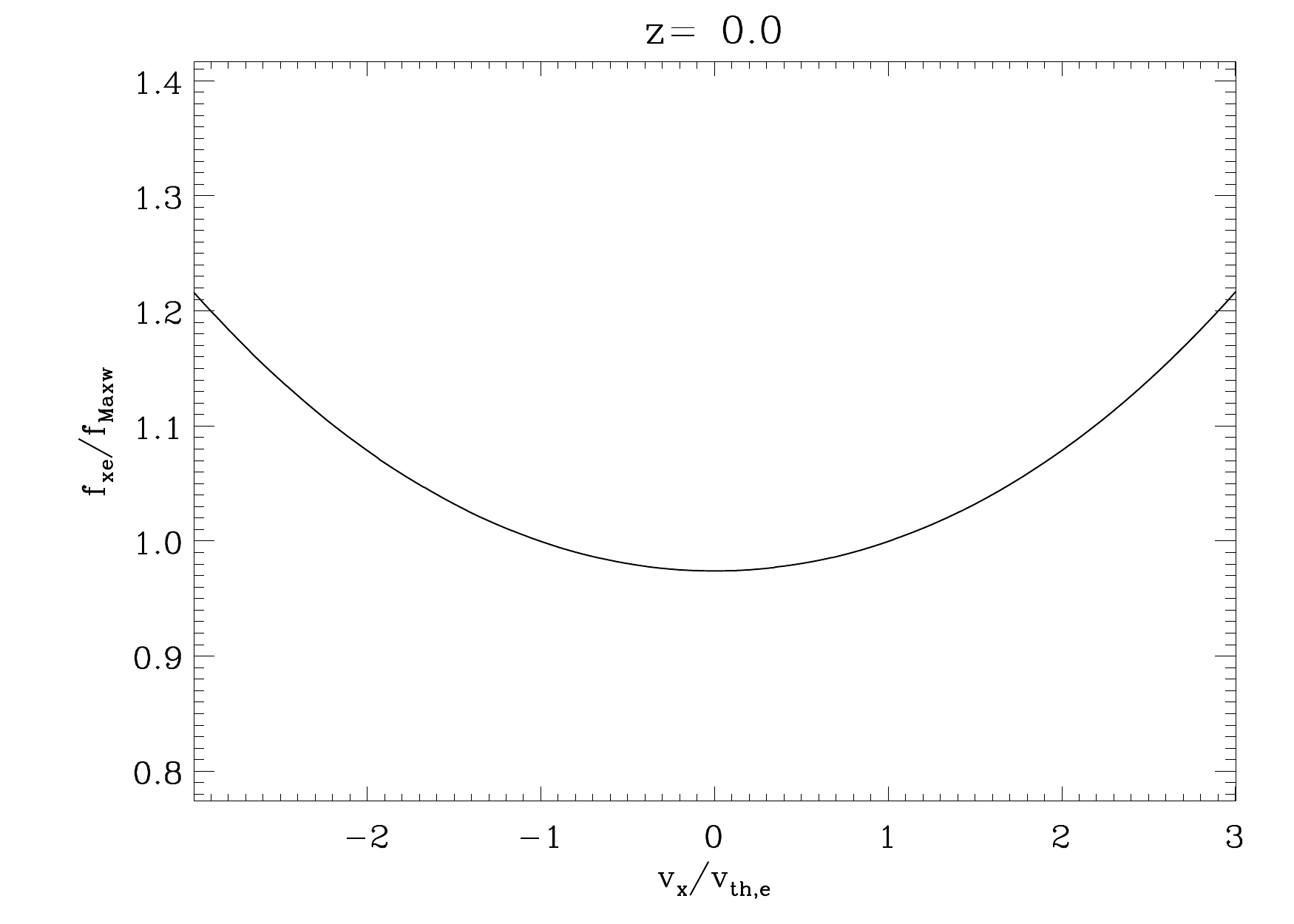}
 \caption{\label{fig:4a}}
\end{subfigure}
\begin{subfigure}[b]{0.5\textwidth}
\includegraphics[width=\textwidth]{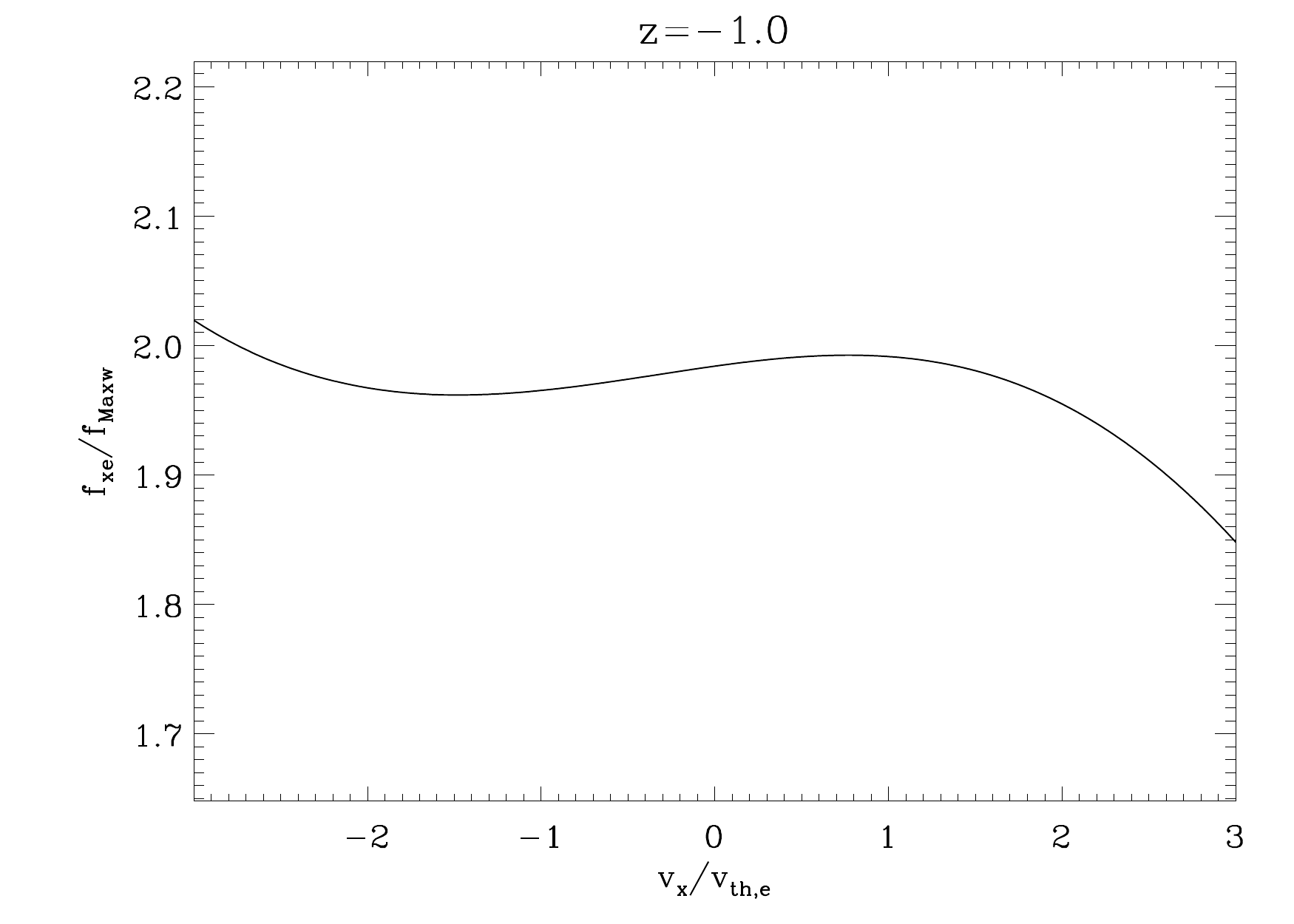}
 \caption{\label{fig:4b}}
\end{subfigure}
\begin{subfigure}[b]{0.5\textwidth}
\includegraphics[width=\textwidth]{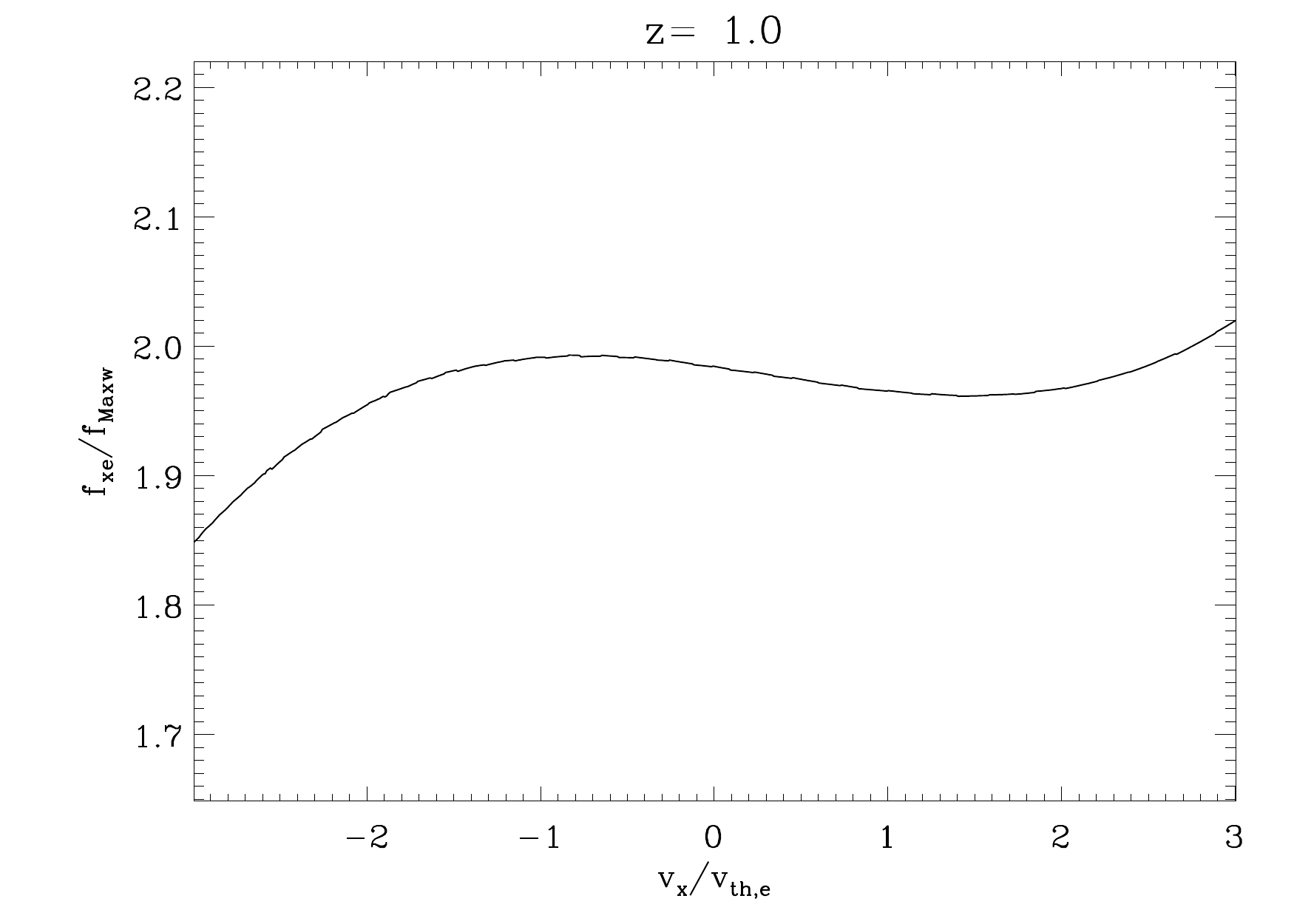}
 \caption{\label{fig:4c}}
\end{subfigure}
\caption{The $v_x$ variation of $f_e/f_{Maxw,e}$ for $z/L=0$ (\ref{fig:4a}),  $z/L=-1$ (\ref{fig:4b}) and  $z/L=1$ (\ref{fig:4c}). $\beta_{pl}=0.85$ and $\delta_e=0.15$. Note the antisymmetry of the $z=\pm 1$ plots with respect to each other.}
 \end{figure}

\begin{figure}
\centering
\begin{subfigure}[b]{0.5\textwidth}
\includegraphics[width=\textwidth]{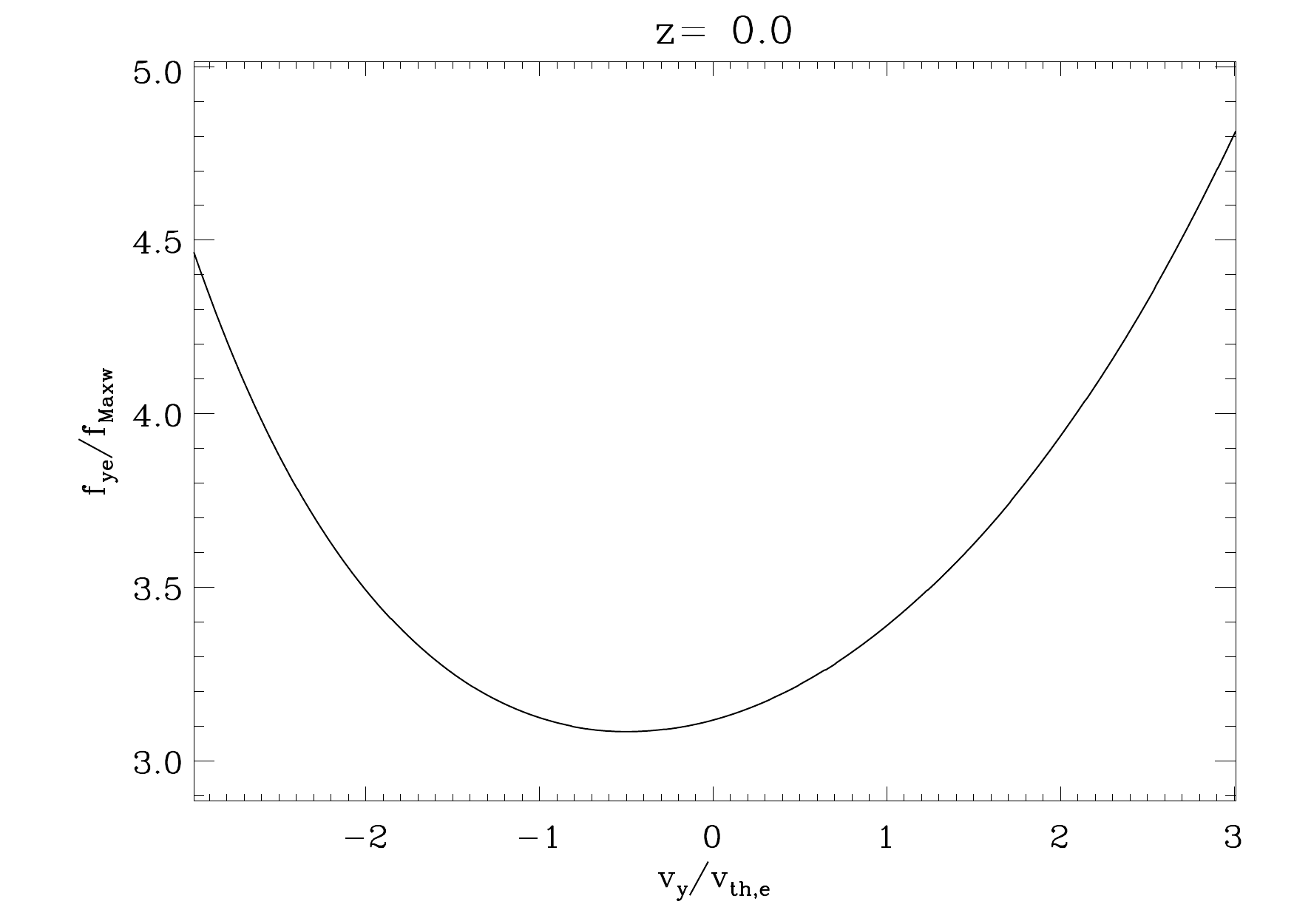}
 \caption{\label{fig:5a}}
\end{subfigure}
\begin{subfigure}[b]{0.5\textwidth}
\includegraphics[width=\textwidth]{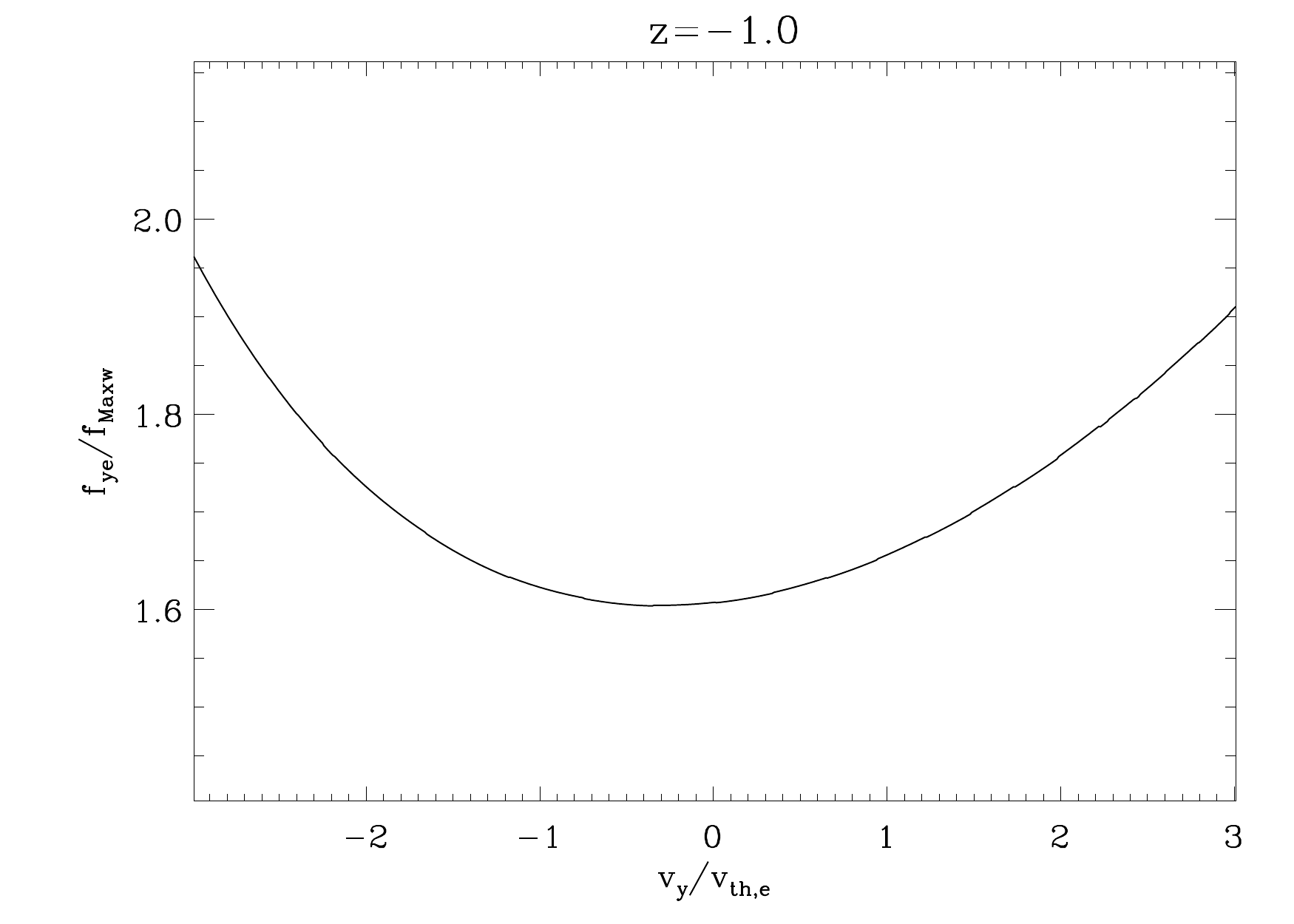}
 \caption{\label{fig:5b}}
\end{subfigure}
\begin{subfigure}[b]{0.5\textwidth}
\includegraphics[width=\textwidth]{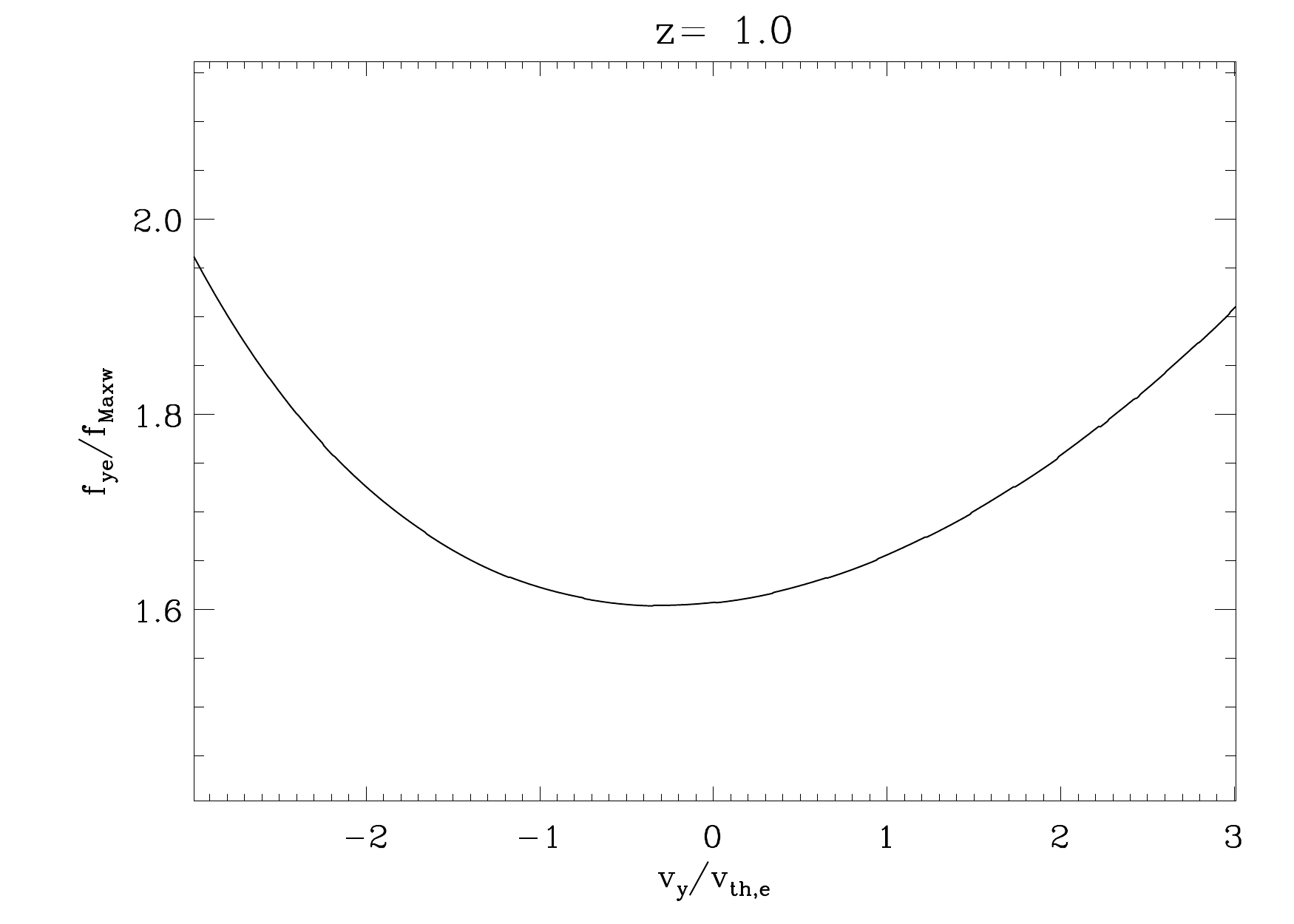}
 \caption{\label{fig:5c}}
\end{subfigure}
\caption{The $v_y$ variation of $f_e/f_{Maxw,e}$ for $z/L=0$ (\ref{fig:5a}),  $z/L=-1$ (\ref{fig:5b}) and  $z/L=1$ (\ref{fig:5c}). $\beta_{pl}=0.85$ and $\delta_e=0.15$. Note the symmetry of the $z=\pm 1$ plots with respect to each other.}
 \end{figure}

A first look at the plots also seems to indicate that the shape of the DF resembles the shape of a Maxwellian. Motivated by this similarity, we define a Maxwellian DF by
\begin{equation}
f_{Maxw,s}=\frac{n_0}{(\sqrt{2\pi}v_{th,s})^3}\exp\left[\frac{\left(\mathbf{v}-\langle\mathbf{v}\rangle_s(z)\right)^2}{2v_{th,s}^2}\right]. \label{eq:Maxshift}
\end{equation}   
The Maxwellian distribution reproduces the same first order moment in terms of $z$ as the equilibrium solution does, namely $n_0\langle\mathbf{v}\rangle_s$, and a spatially uniform number density, namely $n_0$. However it is not a solution of the Vlasov equation and hence not an equilibrium solution. See ref. \cite{Hesse-2005} for an example of Particle in Cell simulations with a force-free field, initiated with a distribution of this type. To highlight the difference between the two distribution functions, we plot the both the $v_x$ and $v_y$ variation of the ratio of the DF, with the Maxwellian of equation (\ref{eq:Maxshift}) for both ions and electrons in figures (\ref{fig:2a})-(\ref{fig:5c}). As we can see, in all plots the ratio deviates from unity, and in some cases these deviations are substantial. This shows that the inital impression is somewhat misleading. We also observe a symmetry in that the $v_y$ dependent plots are even in $z$, since $A_y$ and $\langle v_y\rangle _s$ are even in $z$.

\begin{figure}
\centering
\begin{subfigure}[b]{0.5\textwidth}
\includegraphics[width=\textwidth]{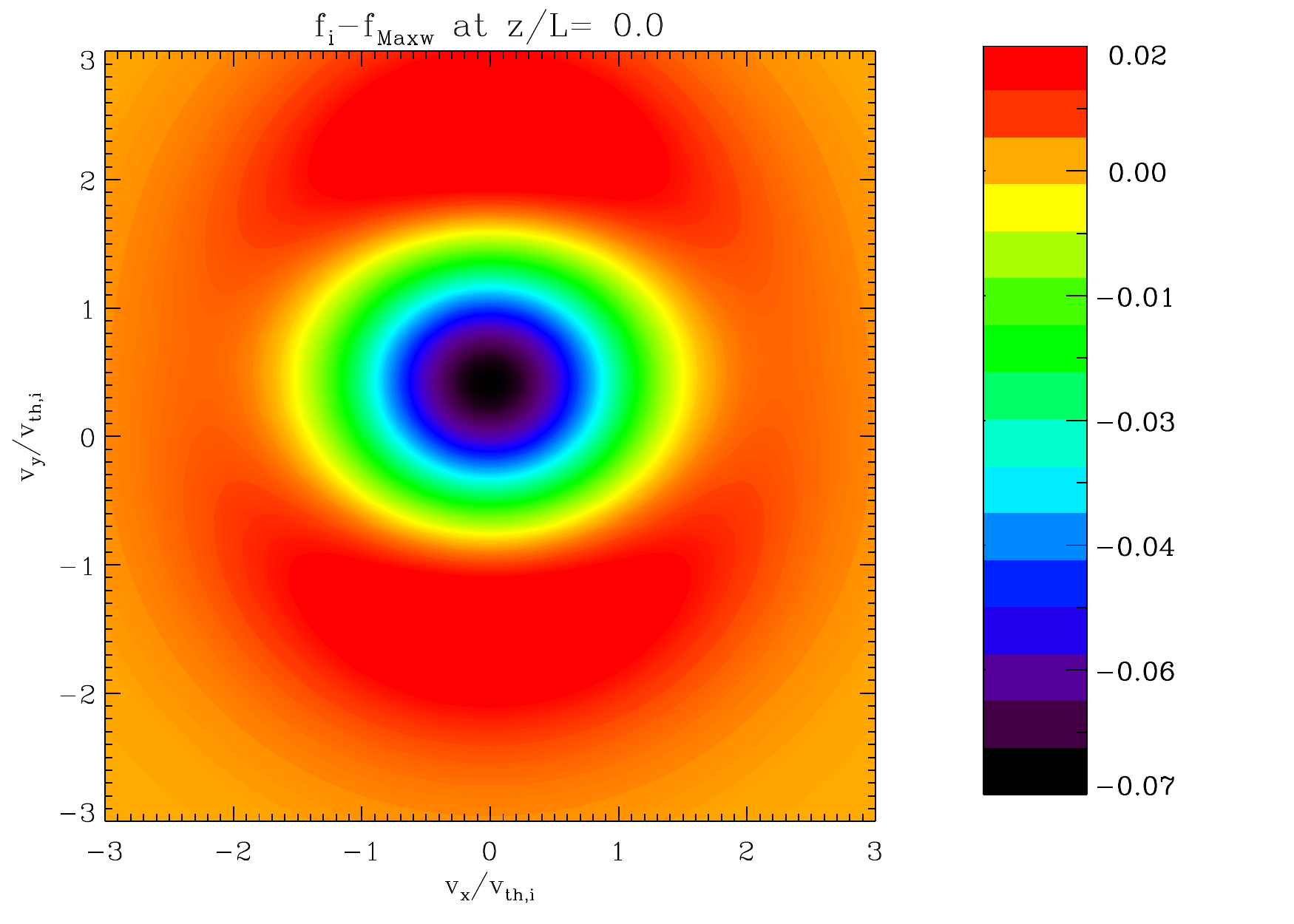}
 \caption{\label{fig:6a}}
\end{subfigure}
\begin{subfigure}[b]{0.5\textwidth}
\includegraphics[width=\textwidth]{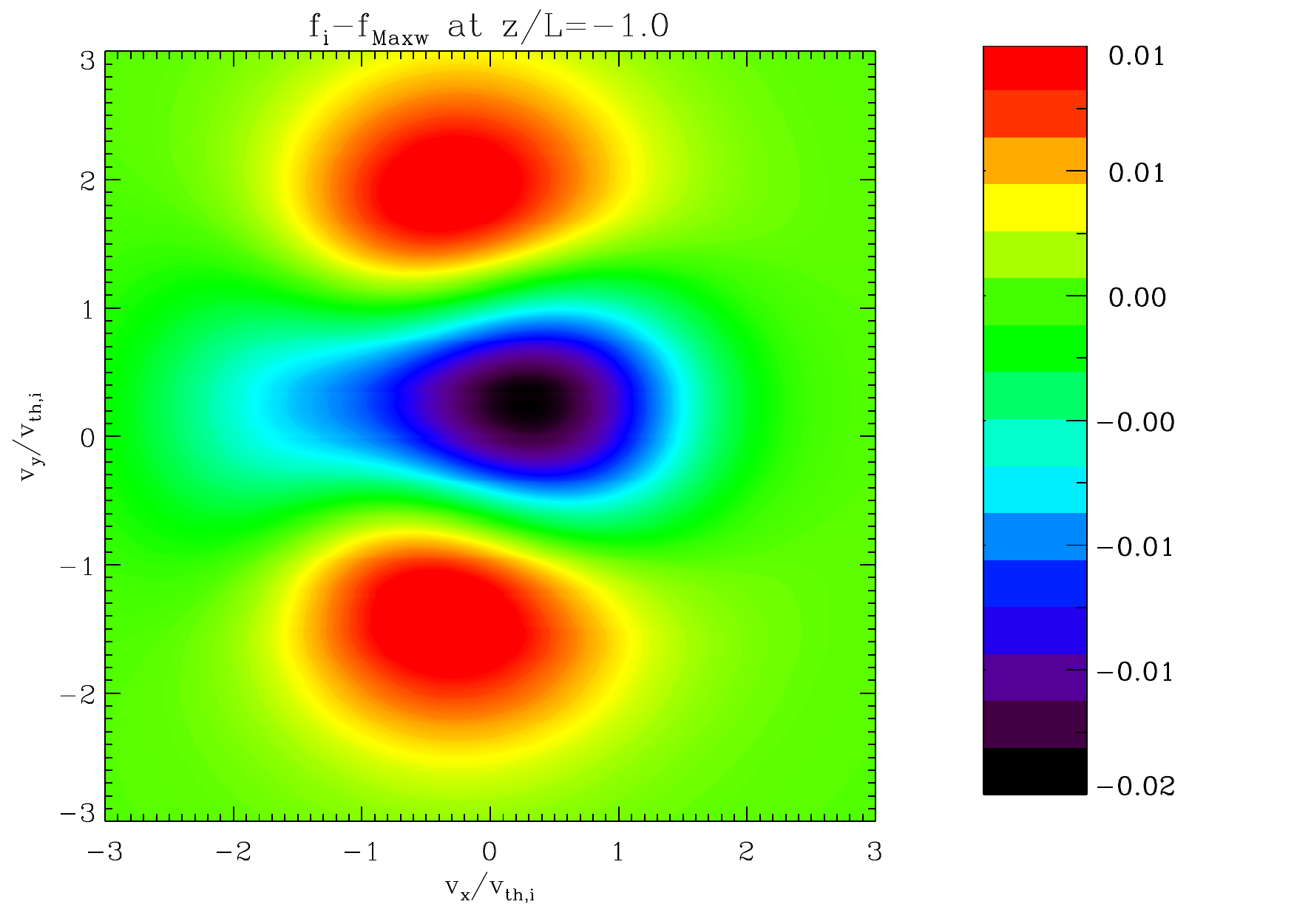}
 \caption{\label{fig:6b}}
\end{subfigure}
\begin{subfigure}[b]{0.5\textwidth}
\includegraphics[width=\textwidth]{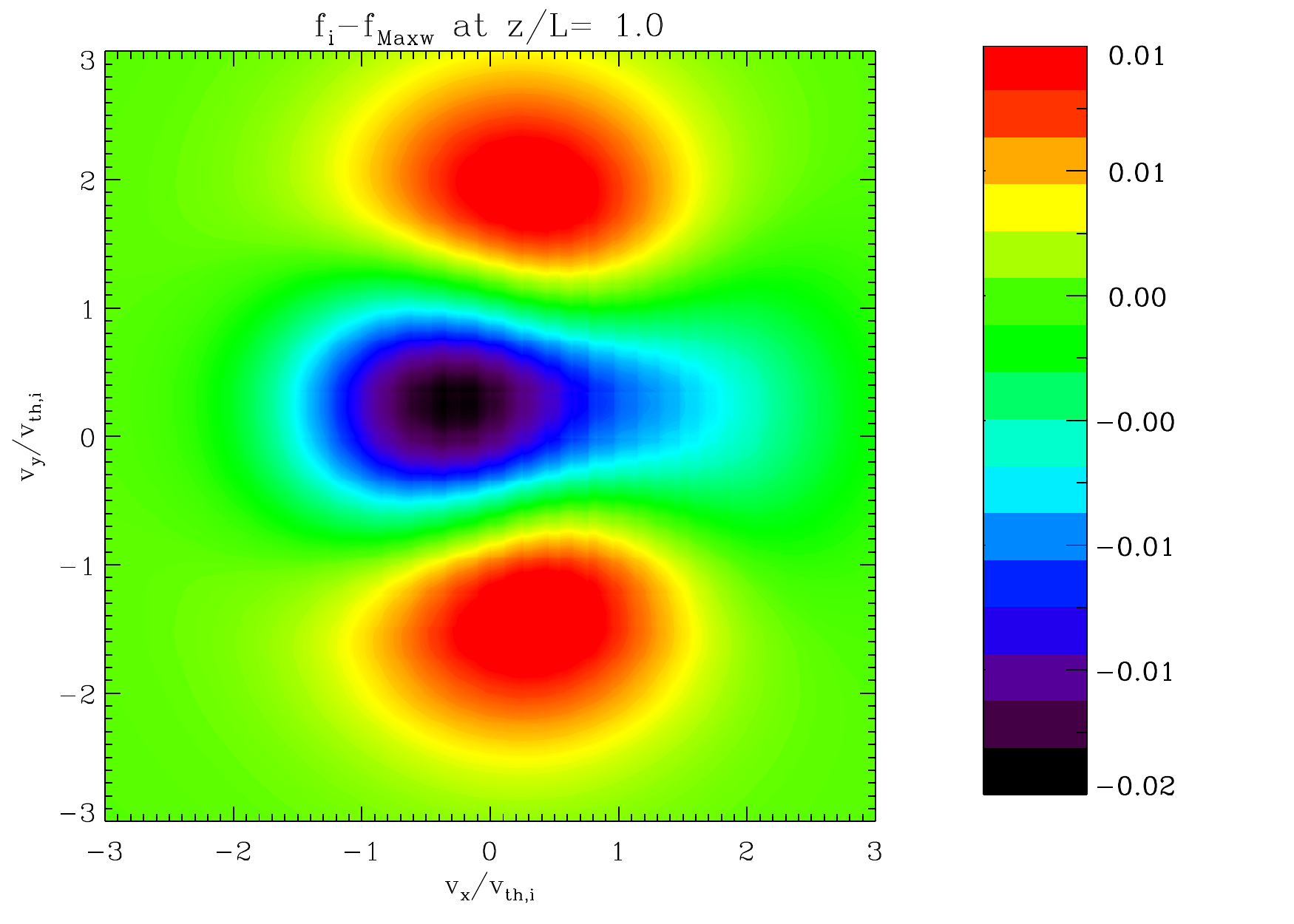}
 \caption{\label{fig:6c}}
\end{subfigure}
\caption{Contour plots of $f_i-f_{Maxw,i}$ for $z/L=0$ (\ref{fig:6a}),  $z/L=-1$ (\ref{fig:6b}) and  $z/L=1$ (\ref{fig:6c}). $\beta_{pl}=0.85$ and $\delta_i=0.15$. Note the antisymmetry of the $z=\pm 1$ plots with respect to each other.}
 \end{figure}

\begin{figure}
\centering
\begin{subfigure}[b]{0.5\textwidth}
\includegraphics[width=\textwidth]{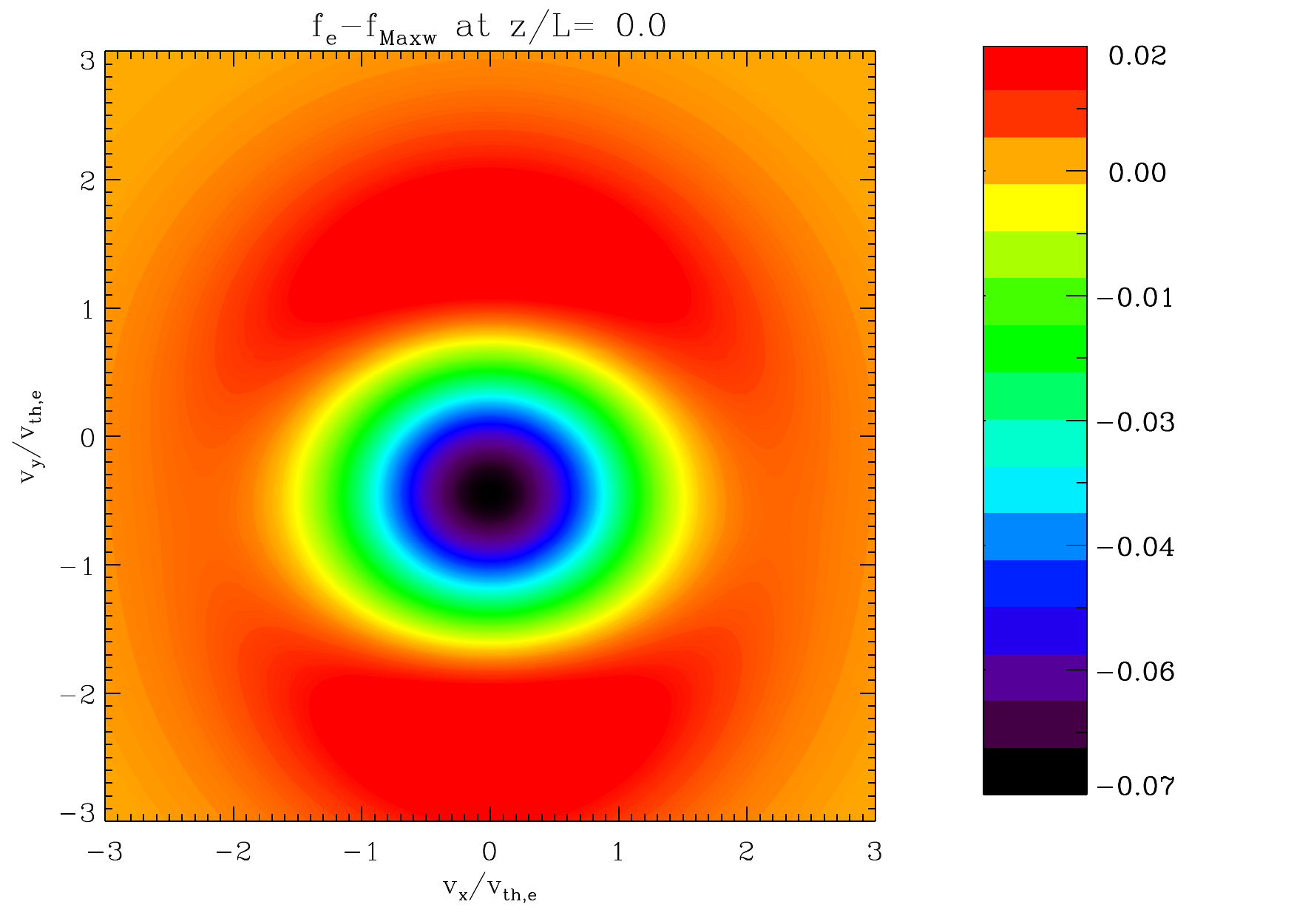}
 \caption{\label{fig:7a}}
\end{subfigure}
\begin{subfigure}[b]{0.5\textwidth}
\includegraphics[width=\textwidth]{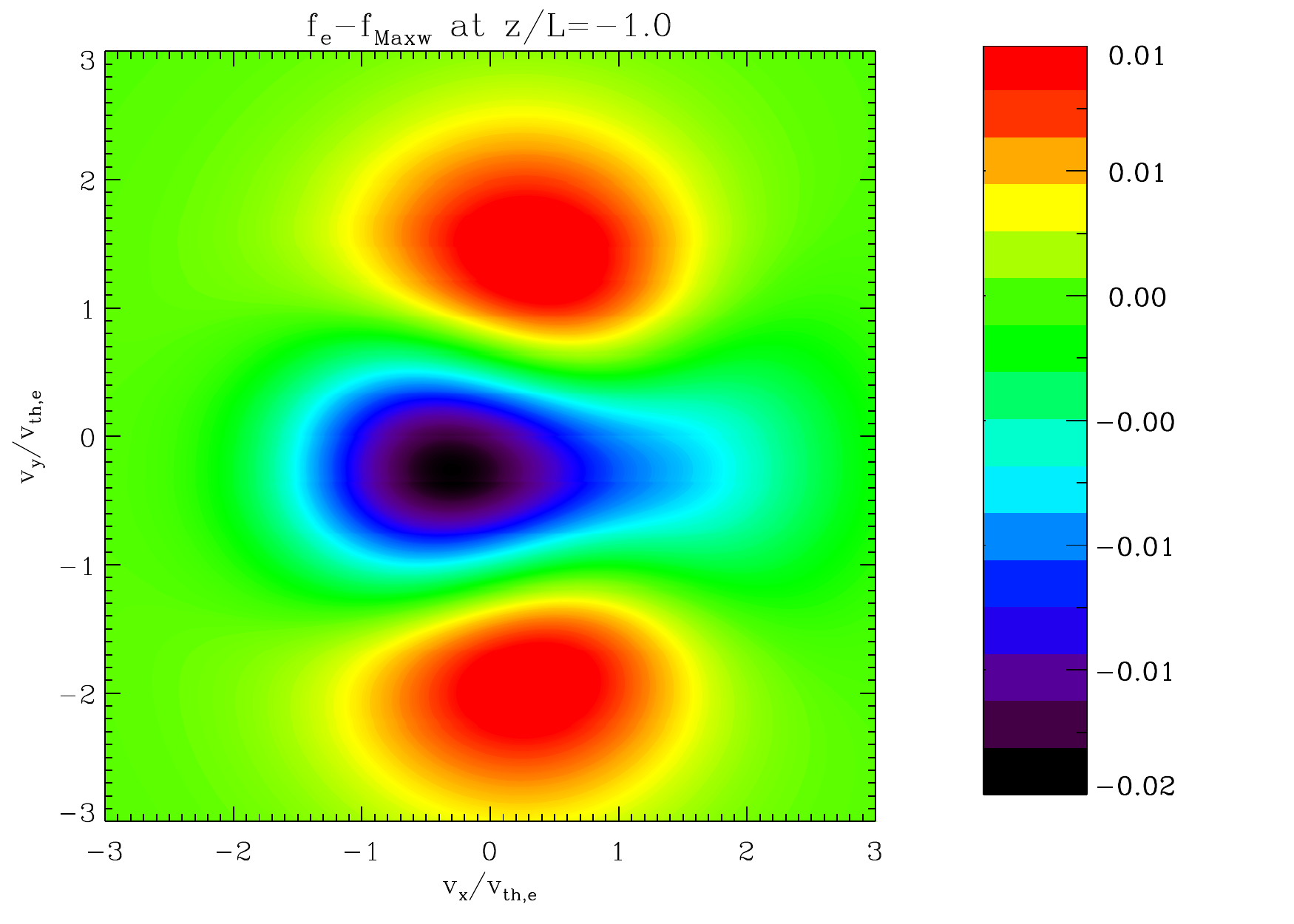}
 \caption{\label{fig:7b}}
\end{subfigure}
\begin{subfigure}[b]{0.5\textwidth}
\includegraphics[width=\textwidth]{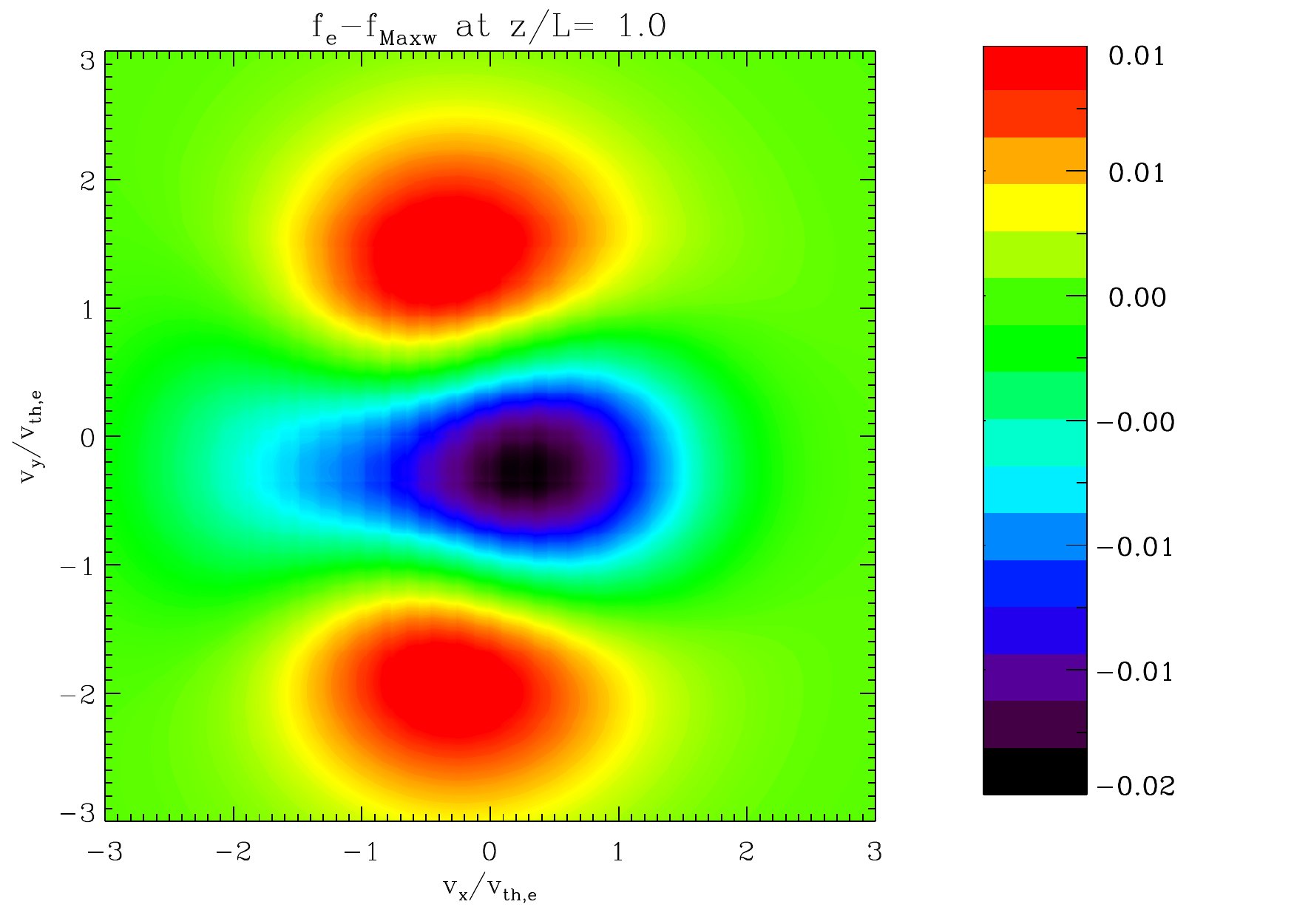}
 \caption{\label{fig:7c}}
\end{subfigure}
\caption{Contour plots of $f_e-f_{Maxw,e}$ for $z/L=0$ (\ref{fig:7a}),  $z/L=-1$ (\ref{fig:7b}) and  $z/L=1$ (\ref{fig:7c}). $\beta_{pl}=0.85$ and $\delta_e=0.15$. Note the antisymmetry of the $z=\pm 1$ plots with respect to each other.}
 \end{figure}

To further see the deviations of $f_s$ from the Maxwellian, we present contour plots of the difference $f_s-f_{Maxw,s}$ in figures (\ref{fig:6a})-(\ref{fig:7c}) over $(v_x,v_y)$ space for various $z$ values. One observation we can make from these is that there is a certain symmetry with respect to both velocity direction and the value of $z$. For example it seems that $f_s$ is symmetric under the transformation $(v_x\to -v_x, \,z\to -z)$. This seems reasonable since $A_x$ is dynamically equivalent to an odd function of $z$, by a gauge transformation, as $B_y$ is even. For a plasma-beta modestly below unity, and thermal Larmor radius roughly 15$\%$ of the current sheet width, we find distributions that are roughly Maxwellian in shape, but `shallower' at the centre of the sheet. At the outer edges of the sheet, this shallowness assumes a drop-shaped depression in the $v_x$  direction, with localised differences for large $v_y$.

\section{Summary \& conclusions}
This paper contains a first analysis of a DF capable of describing low plasma beta, nonlinear force-free collisionless equilibria. In this paper, the emphasis has been on a discussion of properties of the DF. By using expressions for the moments of the DFs we have derived the relationships between the micro- and macroscopic parameters of the equilibrium, in particular the current sheet width. We have presented line-plots of the electron DF in the $v_x$ direction as a representative example. These show that the DF has a single maximum in the $v_x$ direction, and \emph{seems} to resemble a Maxwellian, at least for the parameter range studied. However, a detailed comparison with a Maxwellian describing the same particle density and average velocity/current density shows that there are significant deviations. This was corroborated by contour plots of the difference between the DF and the Maxwellian in the $(v_x,v_y)$ plane. 

While it has been shown \cite{Allanson-2015a} that the infinite series over Hermite polynomials are convergent for all parameter values, plotting the DF has been difficult for the low-beta regime, particularly the $v_x$ dependent sum. Hence, further work on attaining numerical convergence for a wider parameter range is essential. It would be particularly interesting to find out whether the DF develops multiple peaks similar to the DF found for an additive form of $P_{zz}$ \cite{Neukirch-2009}.

\section*{Acknowledgments}
The authors gratefully acknowledge the support of the Science and Technology Facilities Council Consolidated Grant ST/K000950/1 (TN \& FW) and a Doctoral Training Grant ST/K502327/1 (OA). We also gratefully acknowledge funding from the Leverhulme Trust F/00268/BB (TN \& FW). The research leading to these results has received funding from the European
Commission’s Seventh Framework Programme FP7 under the grant agreement SHOCK 284515 (OA, TN \& FW). Finally, thanks go to funding from the Engineering and Physical Sciences Research Council Doctoral Training Grant EP/K503162/1 (ST).

\section*{Appendix A: The $v_x$ moment}
The first order moments of the DF are used to calculate the bulk velocity, and in turn the current density. It is useful to calculate the current density from the DF not only as a procedural check, but also to derive relations between the micro- and macroscopic parameters. We now take the first moment of the DF by $v_x$ denoted by $[v_xf_s]$;
\begin{eqnarray*}
[v_xf_s]&=&\displaystyle\frac{1}{m_s^3}\int_{-\infty}^{\infty}\int_{-\infty}^{\infty}\int_{-\infty}^{\infty} v_xf_sd^3p,\\
&=&\displaystyle\frac{n_0e^{-\frac{1}{2\beta_{pl}}}}{(\sqrt{2\pi})m_sv_{th,s}}\sum_{n=0}^\infty b_n\left(\frac{A_y}{B_0L}\right)^n\sum_{m=0}^\infty C_{2m,s}\nonumber\\
&&\times\underbrace{\int_{-\infty}^\infty v_xe^{-\frac{\beta_s}{2m_s}(p_{xs}-q_sA_x)^2}H_{2m}\left(\frac{p_{xs}}{\sqrt{2}m_sv_{th,s}}\right)dp_{xs}}_{I_{v_x}},
\end{eqnarray*}
after both the $p_{ys}$ and $p_{zs}$ integrations. Now, use the Hermite expansion of the exponential \cite{Morse-1953}, to give
\begin{eqnarray}
&&I_{v_x}=\frac{1}{m_s}\int_{-\infty}^\infty (p_{xs}-q_sA_x)H_{2m}\left(\frac{p_{xs}}{\sqrt{2}m_sv_{th,s}}\right)e^{-\frac{\beta_sp_{xs}^2}{2m_s}}\nonumber\\
&&\times\left[\sum_{j=0}^\infty\frac{1}{(j)!}\left(\frac{q_sA_x}{\sqrt{2}m_sv_{th,s}}\right)^jH_{j}\left(\frac{p_{xs}}{\sqrt{2}m_sv_{th,s}}\right)\right]dp_{xs}.
\end{eqnarray}
Define an inner product according to
\begin{equation}
\langle f_1(x),f_2(x)\rangle =\int_{-\infty}^\infty e^{-x^2}f_1(x)f_2(x)dx.
\end{equation}
Then orthogonality of the Hermite polynomials, $\langle H_k(x), H_l(x)\rangle = \sqrt{\pi}2^kk!\delta_{kl}$, and the recurrence relation, $H_{n+1}(x)=2xH_n(x)-2nH_{n-1}(x)$, are used to give
\begin{equation}
\begin{split}
\langle xH_j(x),H_{2m}(x)\rangle&=j\langle H_{j-1}(x),H_{2m}(x)\rangle+\frac{1}{2}\langle H_{j+1}(x),H_{2m}(x)\rangle\\
&=\sqrt{\pi}2^{2m}(2m)!\left(j\delta_{j-1,2m}+\frac{1}{2}\delta_{j+1,2m}\right).
\end{split}
\end{equation}
This allows us to write
\begin{equation}
I_{v_x}=\displaystyle\sqrt{2\pi}v_{th,s}2^{2m}(2m)!\sum_{j=0}^\infty \frac{1}{j!}\left(\frac{q_sA_x}{\sqrt{2}m_sv_{th,s}}\right)^j\biggr[\sqrt{2}m_sv_{th,s}\left(j\delta_{j-1,2m}+\frac{1}{2}\delta_{j+1,2m}\right)-q_sA_x\delta_{j,2m}\biggr].\nonumber
\end{equation}
So we have
\begin{eqnarray*}
[v_xf_s]&=&\displaystyle\frac{n_0e^{-\frac{1}{2\beta_{pl}}}}{m_s}\sum_{n=0}^\infty b_n\left(\frac{A_y}{B_0L}\right)^n\sum_{m=0}^\infty C_{2m,s}2^{2m}(2m)!\\
&&\times\sum_{j=0}^\infty \frac{1}{j!}\left(\frac{q_sA_x}{\sqrt{2}m_sv_{th,s}}\right)^j\biggr[\sqrt{2}m_sv_{th,s}\left(j\delta_{j-1,2m}+\frac{1}{2}\delta_{j+1,2m}\right)-q_sA_x\delta_{j,2m}\biggr].
\end{eqnarray*}
reducing to
\begin{eqnarray}
\displaystyle[v_xf_s]&=&\displaystyle\left(\frac{m_sv_{th,s}^2}{q_sB_0L}\right)n_0e^{-\frac{1}{2\beta_{pl}}}\sum_{n=0}^\infty b_n\left(\frac{A_y}{B_0L}\right)^n\sum_{m=1}^\infty a_{2m}2m\left(\frac{A_x}{B_0L}\right)^{2m-1}\nonumber \\
&=&\left(\frac{m_sv_{th,s}^2}{q_sP_0}\right)n_0\frac{\partial \bar{P}_{zz}}{\partial A_x}=\frac{\beta_e\beta_i}{\beta_e+\beta_i}\left(\frac{1}{q_s\beta_s}\right)\frac{\partial \bar{P}_{zz}}{\partial A_x}\label{eq:verify}
\end{eqnarray}

The $x$ component of current density is defined as $j_x=\sum_sq_s[v_xf_s]$, giving
\begin{eqnarray}
j_x&=&\frac{\beta_e\beta_i}{\beta_e+\beta_i}\frac{\partial \bar{P}_{zz}}{\partial A_x}\sum_s\frac{1}{\beta_s}=\frac{\partial \bar{P}_{zz}}{\partial A_x}\nonumber\\
\implies j_x&=&\frac{\partial \bar{P}_{zz}}{\partial A_x},
\end{eqnarray}
reproducing the familiar result \cite{Channell-1976, Harrison-2009a, Schindlerbook, Mynick-1979a}.
The first moment of the DF can also be used to calculate the bulk velocity in terms of the microscopic parameters;
\begin{equation}
\langle v_x\rangle_s=\frac{[v_xf_s]}{n_s}=\frac{j_x}{q_s\beta_sP_0},
\end{equation}
using equation (\ref{eq:verify}). Then, by using the current density for the FFHS,
\begin{equation}
\mathbf{j}=\frac{B_0}{\mu_0L}\left(\frac{\text{sinh}\left(\frac{z}{L}\right)}{\text{cosh}^2\left(\frac{z}{L}\right)},\frac{1}{\text{cosh}^2\left(\frac{z}{L}\right)}   ,0 \right),
\end{equation}
we have the fluid flow in $x$
\begin{equation}
\langle v_x\rangle_s=\frac{B_0}{\mu_0 Lq_s\beta_sP_0}\frac{\text{sinh}\left(\frac{z}{L}\right)}{\text{cosh}^2\left(\frac{z}{L}\right)}.
\end{equation}

\section*{Appendix B: The $v_y$ moment}
By a completely analogous calculation, we derive the $v_y$ moment of the DF,
\begin{eqnarray*} 
[v_yf_s]&=&\displaystyle\left(\frac{m_sv_{th,s}^2}{P_0q_s}\right)n_0\frac{\partial \bar{P}_{zz}}{\partial A_y}\\
&=&\displaystyle\frac{\beta_e\beta_i}{\beta_e+\beta_i}\left(\frac{m_sv_{th,s}^2}{q_s}\right)\frac{\partial \bar{P}_{zz}}{\partial A_y}
\end{eqnarray*}
Again, the current density $j_y=\sum_sq_s[v_yf_s]$ gives
\begin{eqnarray*}
j_y&=&\displaystyle\frac{\beta_e\beta_i}{\beta_e+\beta_i}\frac{\partial P_{zz}}{\partial A_y}\sum_sm_sv_{th,s}^2=\frac{\partial \bar{P}_{zz}}{\partial A_y}\\
\implies j_y&=&\frac{\partial \bar{P}_{zz}}{\partial A_y}.
\end{eqnarray*}
We can also calculate the bulk velocity in terms of the microscopic parameters;
\begin{equation}
\langle v_y\rangle_s=\frac{B_0}{\mu_0 Lq_s\beta_sP_0}\frac{1}{\text{cosh}^2\left(\frac{z}{L}\right)}.
\end{equation}

%\newpage

%\bibliographystyle{plain}
%\bibliography{/Users/Oliver/Dropbox/biblio}                  %Home

\newpage

\newpage

\newpage

\newpage

\newpage

\newpage

\newpage

%bibliographystyle{plain}

%\bibliography{/user/oliver/Dropbox/biblio}   %Maths
%\bibliography{/Users/Oliver/Dropbox/biblio}                  %Home

\begin{thebibliography}{10}

\bibitem{Abraham-Shrauner-2013}
B.~{Abraham-Shrauner}.
\newblock {Force-free Jacobian equilibria for Vlasov-Maxwell plasmas}.
\newblock {\em Physics of Plasmas}, 20(10):102117, October 2013.

\bibitem{Allanson-2015a}
O.~{Allanson}, T.~{Neukirch}, F.~{Wilson}, and S.~{Troscheit}.
\newblock {An exact collisionless equilibrium for the Force-Free Harris Sheet
  with low plasma beta}.
\newblock {\em Physics of Plasmas}, 22(11), November 2015.

\bibitem{Arfkenbook}
George~B. Arfken and Hans~J. Weber.
\newblock {\em Mathematical methods for physicists}.
\newblock Harcourt/Academic Press, Burlington, MA, fifth edition, 2001.

\bibitem{Artemyev-2011}
A.~V. {Artemyev}.
\newblock {A model of one-dimensional current sheet with parallel currents and
  normal component of magnetic field}.
\newblock {\em Physics of Plasmas}, 18(2):022104, February 2011.

\bibitem{Attico-1999}
N.~{Attico} and F.~{Pegoraro}.
\newblock {Periodic equilibria of the Vlasov-Maxwell system}.
\newblock {\em Physics of Plasmas}, 6:767--770, March 1999.

\bibitem{Bell-1934}
E.~T. Bell.
\newblock Exponential polynomials.
\newblock {\em Ann. of Math. (2)}, 35(2):258--277, 1934.

\bibitem{Bertotti-1963}
B.~{Bertotti}.
\newblock {Fine structure in current sheaths}.
\newblock {\em Annals of Physics}, 25:271--289, December 1963.

\bibitem{Bilodeau-1962}
G.~G Bilodeau.
\newblock The weierstrass transform and hermite polynomials.
\newblock {\em Duke Mathematical Journal}, 29(2):293--308, 1962.

\bibitem{Bobrova-2001}
N.~A. {Bobrova}, S.~V. {Bulanov}, J.~I. {Sakai}, and D.~{Sugiyama}.
\newblock {Force-free equilibria and reconnection of the magnetic field lines
  in collisionless plasma configurations}.
\newblock {\em Physics of Plasmas}, 8:759--768, March 2001.

\bibitem{Bobrova-2003}
N.~A. {Bobrova}, S.~V. {Bulanov}, G.~E. {Vekstein}, J.-I. {Sakai},
  K.~{Machida}, and T.~{Haruki}.
\newblock {Tearing instability of a force-free magnetic configuration in a
  collisionless plasma}.
\newblock {\em Plasma Physics Reports}, 29:449--458, June 2003.

\bibitem{Bobrova-1979}
N.~A. {Bobrova} and S.~I. {Syrovatski{\v i}}.
\newblock {Violent instability of one-dimensional forceless magnetic field in a
  rarefied plasma}.
\newblock {\em Soviet Journal of Experimental and Theoretical Physics Letters},
  30:535--+, November 1979.

\bibitem{Channell-1976}
P.~J. {Channell}.
\newblock {Exact Vlasov-Maxwell equilibria with sheared magnetic fields}.
\newblock {\em Physics of Fluids}, 19:1541--1545, October 1976.

\bibitem{Correa-Restrepo-1993}
D.~{Correa-Restrepo} and D.~{Pfirsch}.
\newblock {Negative-energy waves in an inhomogeneous force-free Vlasov plasma
  with sheared magnetic field}.
\newblock {\em Physical Review E}, 47:545--563, January 1993.

\bibitem{Fitzpatrick-2007}
R.~{Fitzpatrick}.
\newblock {Interaction of scrape-off layer currents with magnetohydrodynamical
  instabilities in tokamak plasmas}.
\newblock {\em Physics of Plasmas}, 14(6):062505, June 2007.

\bibitem{Fitzpatrickbook}
R.~Fitzpatrick.
\newblock {\em Plasma Physics: An Introduction}.
\newblock CRC Press, Taylor \& Francis Group, 2014.

\bibitem{Grad-1961}
H.~{Grad}.
\newblock {Boundary Layer between a Plasma and a Magnetic Field}.
\newblock {\em Physics of Fluids}, 4:1366--1375, November 1961.

\bibitem{Harrison-2009b}
M.~G. {Harrison} and T.~{Neukirch}.
\newblock {One-Dimensional Vlasov-Maxwell Equilibrium for the Force-Free Harris
  Sheet}.
\newblock {\em Physical Review Letters}, 102(13):135003--+, April 2009.

\bibitem{Harrison-2009a}
M.~G. {Harrison} and T.~{Neukirch}.
\newblock {Some remarks on one-dimensional force-free Vlasov-Maxwell
  equilibria}.
\newblock {\em Physics of Plasmas}, 16(2):022106--+, February 2009.

\bibitem{Hesse-2005}
M.~{Hesse}, M.~{Kuznetsova}, K.~{Schindler}, and J.~{Birn}.
\newblock {Three-dimensional modeling of electron quasiviscous dissipation in
  guide-field magnetic reconnection}.
\newblock {\em Physics of Plasmas}, 12(10):100704--+, October 2005.

\bibitem{Hewett-1976}
D.~W. {Hewett}, C.~W. {Nielson}, and D.~{Winske}.
\newblock {Vlasov confinement equilibria in one dimension}.
\newblock {\em Physics of Fluids}, 19:443--449, March 1976.

\bibitem{Howes-2006}
G.~G. {Howes}, S.~C. {Cowley}, W.~{Dorland}, G.~W. {Hammett}, E.~{Quataert},
  and A.~A. {Schekochihin}.
\newblock {Astrophysical Gyrokinetics: Basic Equations and Linear Theory}.
\newblock {\em The Astrophysical Journal}, 651:590--614, November 2006.

\bibitem{Johnson-2002}
Warren~P. Johnson.
\newblock The curious history of {F}a\`a di {B}runo's formula.
\newblock {\em Amer. Math. Monthly}, 109(3):217--234, 2002.

\bibitem{Marshbook}
{G.~E.} {Marsh}.
\newblock {\em Force-Free Magnetic Fields: Solutions, Topology and
  Applications}.
\newblock World Scientific, Singapore, 1996.

\bibitem{Morse-1953}
Philip~M. Morse and Herman Feshbach.
\newblock {\em Methods of theoretical physics. 2 volumes}.
\newblock McGraw-Hill Book Co., 1953.

\bibitem{Mynick-1979a}
H.~E. {Mynick}, W.~M. {Sharp}, and A.~N. {Kaufman}.
\newblock {Realistic Vlasov slab equilibria with magnetic shear}.
\newblock {\em Physics of Fluids}, 22:1478--1484, August 1979.

\bibitem{Neukirch-2009}
T.~{Neukirch}, F.~{Wilson}, and M.~G. {Harrison}.
\newblock {A detailed investigation of the properties of a Vlasov-Maxwell
  equilibrium for the force-free Harris sheet}.
\newblock {\em Physics of Plasmas}, 16(12):122102, December 2009.

\bibitem{Nicholson-1963}
R.~B. {Nicholson}.
\newblock {Solution of the Vlasov Equations for a Plasma in an Externally
  Uniform Magnetic Field}.
\newblock {\em Physics of Fluids}, 6:1581--1586, November 1963.

\bibitem{Priest-2000}
E.~{Priest} and T.~{Forbes}.
\newblock {\em {Magnetic Reconnection}}.
\newblock Cambridge University Press, Cambridge, UK, June 2000.

\bibitem{Sakurai-1989}
T.~{Sakurai}.
\newblock {Computational modeling of magnetic fields in solar active regions}.
\newblock {\em Space Science Reviews}, 51:11--48, October 1989.

\bibitem{Schindlerbook}
K.~Schindler.
\newblock {\em Physics of Space Plasma Activity}.
\newblock Cambridge University Press, Cambridge, UK, 2007.

\bibitem{Sestero-1966}
A.~{Sestero}.
\newblock {Vlasov Equation Study of Plasma Motion across Magnetic Fields}.
\newblock {\em Physics of Fluids}, 9:2006--2013, October 1966.

\bibitem{Sestero-1967}
A.~{Sestero}.
\newblock {Self-Consistent Description of a Warm Stationary Plasma in a
  Uniformly Sheared Magnetic Field}.
\newblock {\em Physics of Fluids}, 10:193--197, January 1967.

\bibitem{Suzuki-2008}
A.~{Suzuki} and T.~{Shigeyama}.
\newblock {A novel method to construct stationary solutions of the
  Vlasov-Maxwell system}.
\newblock {\em Physics of Plasmas}, 15(4):042107--+, April 2008.

\bibitem{Widder-1951}
D.~V. Widder.
\newblock {Necessary and sufficient conditions for the representation of a
  function by a Weierstrass transform}.
\newblock {\em Transactions of the American Mathematical Society}, 71:430--439,
  November 1951.

\bibitem{Widder-1954}
D.~V. Widder.
\newblock The convolution transform.
\newblock {\em Bulletin of the American Mathematical Society}, 60(5):444--456,
  09 1954.

\bibitem{Wiegelmann}
Thomas Wiegelmann and Takashi Sakurai.
\newblock Solar force-free magnetic fields.
\newblock {\em Living Reviews in Solar Physics}, 9(5), 2012.

\bibitem{Wilson-2011}
F.~{Wilson} and T.~{Neukirch}.
\newblock {A family of one-dimensional Vlasov-Maxwell equilibria for the
  force-free Harris sheet}.
\newblock {\em Physics of Plasmas}, 18(8):082108, August 2011.

\end{thebibliography}

\end{document}